\documentclass[lettersize,journal]{IEEEtran}

\usepackage{amsmath,amsfonts}
\usepackage{array}
\usepackage[caption=false,font=normalsize,labelfont=sf,textfont=sf]{subfig}
\usepackage{textcomp}
\usepackage{stfloats}
\usepackage{url}
\usepackage{verbatim}
\usepackage{graphicx}
\usepackage{cite}
\usepackage{caption}
\usepackage{subcaption}
\usepackage{xcolor, color, soul}
\usepackage{xfrac}
\usepackage{soul} 
\usepackage[normalem]{ulem} 
\usepackage{amsthm}

\usepackage{etoolbox} 

\usepackage{subcaption}  
\usepackage{graphicx} 
\usepackage[utf8]{inputenc}
\PassOptionsToPackage{caption=false}{subfig}
\usepackage{tabularx} 
\usepackage{booktabs} 
\sethlcolor{white} 
\usepackage{amssymb}
\usepackage{algorithm}
\usepackage{algpseudocode}
\usepackage{amsmath}
\usepackage{multirow}
\usepackage{xcolor}
\definecolor{myblue}{RGB}{0,0,0}

\newcounter{subsubsubsection}[subsubsection]
\renewcommand{\thesubsubsubsection}{\thesubsubsection.\arabic{subsubsubsection}}
\newcommand{\subsubsubsection}[1]{%
  \refstepcounter{subsubsubsection}%
  \paragraph{\thesubsubsubsection\quad #1}%
}

\usepackage{etoolbox}
\makeatletter
\pretocmd{\subsubsubsection}{\addcontentsline{toc}{paragraph}{\protect\numberline{\thesubsubsubsection}#1}}{}{}
\makeatother

\begin{document}

\title{ RSMA‑Enabled Hierarchical UAV Networks with Non-Linear Energy Harvesting: Outage Probability Analysis and UAV Placement Optimization}

\author{Faicel Khennoufa, Khelil Abdellatif, Metin Ozturk,~\IEEEmembership{ Senior Member,~IEEE}, Halim Yanikomeroglu,~\IEEEmembership{Fellow,~IEEE}, Safwan Alfattani,~\IEEEmembership{ Member,~IEEE}
\thanks{ F. Khennoufa is with Innovative Technologies Laboratory (LTI), Ecole Nationale Supérieure des Technologies Avancées (ENSTA), Department of Basic Training in Computer Science, Algiers, Algeria, email: faicel.khennoufa@ensta.edu.dz. A. Khelil is with LGEERE Laboratory, Department of Electrical Engineering, Echahid Hamma Lakhdar University, El-Oued, Algeria, email: abdellatif-khelil@univ-eloued.dz. M. Ozturk is with Electrical and Electronics Engineering, Ankara Yıldırım Beyazıt University, Ankara, Turkiye, email: metin.ozturk@aybu.edu.tr. H. Yanikomeroglu is with Non-Terrestrial Network (NTN) Laboratory, Department of Systems and Computer Engineering, Carleton University, Ottawa, K1S 5B6, ON, Canada, email: halim@sce.carleton.ca. S. Alfattani is with King AbdulAziz University, Saudi Arabia, email: smalfattani@kau.edu.sa.

This work was supported in part by the Natural Sciences and Engineering Research Council of Canada (NSERC) Discovery Grant; in part by Deanship of Scientific Research (DSR), King Abdulaziz University, Jeddah, Saudi Arabia, under the Indexed Publication Program (IPP); in part by the Ministry of Higher Education and Scientific Research of Algeria (MESRS); and in part by the Scientific and Technological Research Council of Türkiye (TÜBİTAK).

}
}

\markboth{Accepted in IEEE Transactions on Vehicular Technology 
}%
{Shell \MakeLowercase{\textit{et al.}}: A Sample Article Using IEEEtran.cls for IEEE Journals}


\maketitle
\begin{abstract}
Uncrewed aerial vehicles (UAVs) are expected to enhance connectivity, extend network coverage, and support advanced communication services in sixth-generation (6G) cellular networks, particularly in public and civil applications. Although multi-UAV systems offer greater efficiency and cost-effectiveness than single-UAV deployments, their implementation still faces several fundamental challenges that limit their reliability, sustainability, and scalability.
The limited onboard energy restricts mission duration and communication continuity. Therefore, wireless energy harvesting (EH) emerges as a promising solution to overcome this limitation. However, terrestrial energy sources experience path loss, making EH from surrounding UAVs more sustainable. Moreover, rate-splitting multiple access (RSMA) remains insufficiently explored in hierarchical UAV networks under hardware impairments (HWI) and imperfect channel state information (ICSI).
This paper proposes a hierarchical ad hoc UAV network with non-linear EH and RSMA to enhance both energy and cost efficiency, where UAVs harvest energy from surrounding UAVs. For a practical scenario, we consider the effect of HWI and ICSI in our proposed system. 
To the best of the authors’ knowledge, this study is the first to investigate such a scenario in the literature. The outage probability expressions for ground Internet of things (IoT) devices, each CMU, and the overall outage probability of the proposed system are derived over Nakagami-$m$ fading channels while considering practical constraints such as HWI, ICSI, and non-linear EH. Additionally, approximate outage probability expressions are derived for high transmit power regimes.
Subsequently, we formulate two optimization problems to enhance reliability and performance. The first jointly optimizes ground IoT device pairing and CMU selection to minimize overall outage probability. The second determines the optimal CMU hovering position and EH factor.
Our findings indicate that the proposed system outperforms all benchmarks in terms of outage probability.

\end{abstract}


\section{Introduction}
\IEEEPARstart{U}{ncrewed} aerial vehicles (UAV) communications have gained significant attention in the deployment of next-generation networks due to their numerous advantages, such as ease of mobility, low cost, {\color{black}rapid on-demand deployment}, and the ability to cover remote or affected areas. These advantages are particularly important in post-disaster rescue operations, where traditional communications outages hinder efficient rescue and relief {\color{black}efforts}~\cite{10500741,10938203}. 
{\color{black}
To this end, ad hoc UAV networks play an important role in delivering temporary, flexible wireless networks that can quickly adapt to changing demands~\cite{10006695}. 
Despite these advances, a major challenge is energy scarcity, which limits operational duration and effectiveness of UAVs, especially in environments where battery replacement or access to traditional energy sources is impractical. To overcome this limitation, wireless energy harvesting (EH) is a green and sustainable technology that can be utilized to extend the network lifetime through scavenging ambient radio frequency (RF) signals to produce usable energy, thereby enhancing their operational reliability during crises~\cite{10806653}. 
}

{\color{black}
Additionally, the dynamic nature of ad hoc or multi-UAV systems presents difficulties in maintaining efficient and interference-resilient communication. Advanced multiple access techniques, such as rate-splitting multiple access (RSMA) are found to be effective solutions compared to non-orthogonal multiple access (NOMA) for improving spectral efficiency and interference management by splitting users' messages into common and private components~\cite{11029408}.
Consequently, the integration of EH and RSMA in ad hoc UAV networks supports the creation of sustainable, efficient, and resilient communication systems for future wireless applications.
}

\subsection{Related Works}
{\color{black}The advancement in multiple access has been toward NOMA, in which users are superposed using shared time-frequency resources~\cite{10430407}. In recent times, RSMA, based on the principle of rate splitting, has been recognized as a promising physical-layer transmission paradigm for non-orthogonal transmission, interference management, and multiple access strategies in next-generation communication systems~\cite{11037391,mao2022rate}. The main idea behind RSMA is to split user messages into common and private parts, with the ability to partially decode interference as well as partially treat interference as noise, as opposed to extreme interference management techniques in NOMA~\cite{mao2022rate}.
Although RSMA has been explored in wireless networks, its application to UAVs in wireless networks has recently attracted significant attention}, inspiring various studies on improving coverage, security, capacity, and reliability \cite{ahmed2025toward}.
{\color{black}The authors in \cite{10411132} focused on maximizing the} energy efficiency and system throughput of an RSMA-enabled UAV system, where a single UAV serves multiple ground users in pairs. 
In \cite{hua2024sum}, the UAV-aided RSMA downlink communication system was proposed, considering user mobility and uniform rectangular array antenna design. They aimed to maximize the sum-rate by jointly optimizing the UAV beamforming matrix, common rate allocation, and UAV trajectory design. 
{\color{black}In \cite{9258414, 10122143}, the authors offered an in-depth study on the integration of RSMA with UAVs downlink communication systems, where the UAV positions and RSMA parameters were jointly optimized to maximize the sum rate, enhancing the resource allocation efficiency. Specifically, the subcarrier allocation was handled via a swap matching algorithm. Their results show that the RSMA schemes outperform orthogonal frequency-division multiple access and approach NOMA performance.}
{\color{myblue}RSMA has also been shown to be more effective in highly dynamic and challenging wireless environments in recent literature. RSMA for covert communications using imperfect channel state information (ICSI) with optimal power allocation was studied in \cite{11441975}. A cooperative RSMA-based integrated sensing and communication system in multi-UAV-based emergency communication systems was proposed in \cite{11370792}, where UAV deployment, user association, and beamforming are jointly optimized. RSMA has also been studied in ultra-reliable communications at finite blocklengths in high mobility vehicular scenarios, which exhibited increased reliability and fairness under highly dynamic channel conditions~\cite{11159579}. These recent advances further motivate the adoption of RSMA in hierarchical UAV networks.}
{\color{black}Although these studies show the benefits of RSMA in a UAV-assisted network, they are mainly focused on single-UAV architectures and ideal system conditions. Hence, the benefits of using RSMA in multi-UAV networks under practical system conditions have not been extensively explored.}
{\color{myblue}In addition, while this paper considers static RSMA resource allocation to enable tractable outage probability analysis, dynamic RSMA optimization and resource allocation in hierarchical UAV systems remain important directions for future research.}

Moreover, multi-UAV systems have gained significant attention due to their ability to coordinate missions and {\color{black}enhance communication efficiency} in smart environments. 
In~\cite{10379154}, a multi-hop UAV relay network was proposed to serve heterogeneous ground users lacking direct links, with optimized resource allocation and deployment to meet diverse user needs and reduce outage probability.
In order to maximize the weighted sum-rate while adhering to the sensing signal-to-noise ratio (SNR) requirement, a coordinated RSMA-based integrated sensing and communication architecture was examined for emergency {\color{black}UAV systems in \cite{yao2024coordinated}. In this system, multiple UAVs simultaneously provided communications to survivors and detected trapped people.}
In \cite{lau2023general}, a UAV swarm network was analyzed regarding outage probability and capacity by characterizing the effects of UAV-to-UAV interference.
In \cite{10840246}, the energy-efficient detection coverage for multi-autonomous aerial vehicles was investigated with joint optimization of moving action and channel access for transmitting data. 
{\color{black}
Despite these advances, most existing multi-UAV studies focus on coordination, deployment, or interference management, while the integration of advanced multiple access schemes, such as RSMA with multi-UAV (e.g., hierarchical UAV) architectures, remains insufficiently investigated.

One promising architecture for multi-UAV systems is the hierarchical UAV framework, where UAVs are organized into functional layers that typically include high-altitude leader UAVs and low-altitude follower UAVs.
}
Upper-layer UAVs handle coordination and communication control, while lower-layer UAVs perform data collection or relaying tasks~\cite{10006695}. {\color{black}A hierarchical UAV network provides} greater coverage and efficiency by delegating tasks between mother and member UAVs, thereby reducing power consumption and latency, and enhancing reliability, flexibility, and scalability in large-scale, mission-critical conditions~\cite{10006695}.
{\color{black}
To overcome the challenges posed by network dynamics, the authors in~\cite{10679214} proposed an intelligent hierarchical UAV slicing framework, demonstrating improved system utility, throughput, and reduced transmission delay compared with benchmark schemes. 
A vision-based formation control approach for outdoor UAV swarms without external positioning devices was presented in~\cite{ma2023vision}, where the hierarchical structure enables the leader UAV to broadcast follower locations for swarm coordination. }
A cooperative framework for automatic dependent surveillance-broadcast was designed in \cite{10930451} for hierarchical UAV networks. In \cite{cui2023distributionally}, hierarchical UAV-assisted multi-access edge computing was optimized to minimize power consumption while maintaining service quality. The authors of \cite{lorincz2021novel} investigated disaster management and real-time monitoring applications by employing hierarchical UAVs along with cloud computing resources for trajectory planning, obstacle avoidance, and object detection in real time. 
{\color{black}However, these works mostly focus on the aspects of networking, control, or computing, and the performance analysis of hierarchical UAV communication systems using advanced multiple access techniques and realistic conditions has not been explored.
}

{\color{black}
On the other hand, energy conservation is a major challenge facing UAVs, especially for long-duration missions, due to limited battery capacity and high energy demands of both flight and wireless communications~\cite{ahmed2025toward}. Therefore, wireless EH has been proposed in the literature as one of the potential solutions to address the energy consumption challenge~\cite{ahmed2025toward,9913422}. 
In \cite{9707780}, UAV-assisted mobile edge computing frameworks with EH and NOMA were investigated, where closed-form expressions for the successful computation and energy consumption probabilities were derived over Nakagami-$m$ fading channels. In \cite{10852192}, joint optimization of task allocation, beamforming, and trajectory was proposed to improve energy efficiency and computation performance.
The outage probability of UAV-assisted SWIPT-MIMO with NOMA systems has been widely studied under practical constraints such as imperfect successive interference cancellation (SIC)~\cite{CAGIRAN2025155713}.
A NOMA-assisted full-duplex cooperative UAV-to-UAV communication system with EH was considered in~\cite{10928333}, where outage probability expressions and cellular links were derived through asymptotic analysis.
In~\cite{10659004}, the authors proposed an interference tolerance-based EH strategy for UAV-assisted intelligent transportation systems to improve energy efficiency and mitigate overlapping interference. The approach utilized an interference hypergraph model and an EH optimization framework with {\color{myblue}ICSI}.
However, most existing works consider linear EH models or traditional UAV architectures, while the impact of non-linear EH models in hierarchical UAV communication systems remains largely unexplored.}




Furthermore, over the last decade, a great amount of effort has been devoted to developing integrated, cost-effective, and energy-efficient RF transceivers. However, these systems suffer from inevitable RF front-end impairments due to component mismatches and manufacturing defects. 
Several studies have investigated UAV-enabled communication systems in the presence of hardware impairments (HWI). 
The authors in \cite{10510457} explored the impact of hardware impairments (HWI) and ICSI on the mean square error of over-the-air computation using UAVs. 
{\color{black}
In \cite{chen2024performance,cheng2025secrecy}, UAV-assisted amplify-and-forward (AF) and decode-and-forward (DF) communication systems with or without NOMA were analyzed in terms of outage probability and secrecy performance. In \cite{10146460}, a UAV-enabled IoT network with RF energy harvesting and DF relaying under HWI was studied in terms of outage probability and energy efficiency.}
In~\cite{10287354}, a deep neural network-based UAV-NOMA system was developed for performance prediction and multi-antenna design under hardware noise, where closed-form outage probability and ergodic capacity were derived.
The authors in \cite{9478941} analyzed the outage probability of a hybrid satellite–terrestrial network, where a satellite–ground user equipment link is aided by multiple AF three-dimensional mobile UAV relays under RF HWI. 


{\color{black}
Despite these important contributions, non-linear EH with advanced multiple access techniques, such as RSMA, under the joint impact of HWI and ICSI in hierarchical UAV communication networks has not been thoroughly investigated in the existing literature. Specifically, the majority of the existing research works on RSMA-based UAV communication systems have considered either single UAV systems or assumed ideal conditions of the systems. This, in fact, does not adequately capture the complexity of multi-UAV deployments. Moreover, most of the existing UAV-assisted EH systems have considered linear EH models, which cannot accurately represent the real-world saturation characteristics of the RF EH circuit. Finally, the impact of both HWI and ICSI has not been considered in the analysis of UAV-assisted wireless systems, even though these factors significantly affect system reliability. Therefore, it is critical to investigate hierarchical UAV-enabled RSMA systems with non-linear EH under the influence of HWI and ICSI. This gap motivates the proposed framework, which integrates RSMA, hierarchical UAV network, practical non-linear EH modeling, RF HWI, and ICSI to provide a more realistic and efficient communication architecture, enabling reliable, energy-efficient, and scalable UAV-assisted wireless networks for future applications.}

\subsection{Contributions}
The increasing use of UAVs in smart environments has made multi-UAV systems an effective way to carry out complex missions \cite{ahmed2025toward}. {\color{black}Among these systems, hierarchical multi-UAV architecture has emerged as an innovative solution for organizing UAVs in functional layers. In this structure, command, coordination, and execution duties are distributed to improve coverage, reduce latency, and maximize resource utilization.} Such architectures are particularly valuable in applications like precision agriculture, disaster monitoring, {\color{black}modern communication networks}, and the rapid deployment of relief activities after disasters, where dedicated UAV networks build temporary communication coverage over wide areas. However, {\color{black}one of the major challenges is the limited battery power of UAVs}, as replacing or charging batteries in real-time is often impractical, especially in mission-critical scenarios or rugged environments. Therefore, proposing solutions to address or mitigate this problem is crucial for the success of these systems. As proposed in previous research~\cite{10806653}, wireless EH technology is a promising and sustainable solution. {\color{black}Although previous research introduced} the idea of harvesting energy from ground sources, for example, in~\cite{ahmed2025toward,9913422,9707780,10852192,10902111,CAGIRAN2025155713,10928333,10659004,10146460}, this method may not be effective in disaster scenarios where the existing communication infrastructure collapses. Moreover, {\color{black}harvesting energy from ground stations is affected by path-loss}, which can reduce the amount of energy to be harvested. {\color{black}In order to address this limitation, wireless EH from surrounding UAVs using their transmitted signals is a viable option.} This solution is best suited for line-of-sight (LoS) environments where unobstructed channels facilitate efficient power transfer, making the network more sustainable and resilient.
{\color{black}Moreover, previous works, such as ~\cite{ahmed2025toward,9913422,9707780,10852192,10902111,CAGIRAN2025155713,10928333,10659004,10146460}, considered linear EH, which is an unrealistic assumption. Numerous experiments on real electromagnetic circuits have demonstrated the non-linearity of their input and output characteristics. }
Therefore, the non-linearity of the EH process is considered in this work. 
In addition, RSMA has received limited attention in the design of multi-UAV network architectures, as most existing works have primarily focused on single-UAV or terrestrial network designs~\cite{10411132,hua2024sum,9258414,10122143}.
{\color{black}In hierarchical UAV networks, multiple relay UAVs simultaneously serve ground users, which introduces strong inter-user interference and heterogeneous channel conditions due to different UAV-ground user distances and mobility patterns. Conventional multiple access techniques (e.g., NOMA) may suffer from performance degradation under such conditions. RSMA has recently emerged as an efficient strategy for interference management by splitting each user’s message into common and private components, enabling partial interference decoding and more flexible resource allocation. Therefore, RSMA is particularly suitable for hierarchical UAV architectures, where efficient interference management and spectrum utilization are critical for reliable multi-user communications.}
Moreover, most prior studies have neglected the negative impact of HWI {\color{black} and ICSI} on UAV network scenarios regardless of the presence of EH and RSMA.

To the best of the authors’ knowledge, hierarchical ad hoc UAV network‑assisted non‑linear EH and RSMA under the impact of HWI {\color{black} and ICSI} have not yet been investigated in the literature.
This work represents the first attempt to address this critical research gap. The system comprises two levels of UAVs: a cluster head UAV (CHU) acting as the source, and cluster member UAVs (CMUs) that harvest energy from the CHU’s transmitted signals and {\color{black}use it to relay information to ground users.} To the best of our knowledge, this is the first study in the open literature to explore such a scenario. {\color{black}We employ RSMA to enhance spectrum efficiency} while considering the effects of HWI {\color{black} and ICSI} on the proposed system under practical EH conditions. Thus, the main contributions of this paper are given as follows:

\begin{itemize}
    \item We propose a novel joint design that integrates non-linear EH from surrounding UAVs’ communication signals with RSMA, offering enhanced spectrum efficiency and sustainable energy management in hierarchical ad hoc UAV networks. {\color{black}To achieve a more realistic evaluation}, we also consider the impact of HWI {\color{black}and ICSI} on the system.
   
    \item We derive the outage probability of the proposed hierarchical ad hoc UAV network that incorporates non-linear EH and RSMA in the presence of HWI {\color{black}and ICSI}. Additionally, we conduct an approximate analysis of the outage probability in the high transmit power regime with HWI {\color{black}and ICSI}. Furthermore, we obtain the outage probability of the system for a single CMU (called CMU outage probability), as well as the overall outage probability of a hierarchical ad hoc UAV network (i.e., outage probability across $L$ CMUs).
   
   \item To demonstrate the superiority of the proposed system, we compare it against two benchmark schemes: an ad hoc hierarchical UAV network using non-linear EH with NOMA, and a ground-based EH system where UAVs harvest energy directly from terrestrial transmitters. The results demonstrate that our system outperforms all benchmarks. 

  \item To address the trade-off between coverage and reliability, we formulate an optimization problem to select the optimal CMU from a set of candidates to minimize the overall system outage probability. To this end, we propose the overall outage probability minimization through joint CMU and IoT selection with channel-aware scheduling (OPM-JCIS) algorithm. Ground IoT device pairing is also considered, in which the devices with the strongest channel conditions are selected within each CMU to maximize the overall outage probability performance. The proposed OPM-JCIS algorithm is compared with the case where all CMUs are active and the random selection ({\color{black}RSEL}) method. The results demonstrate that the proposed algorithm achieves higher performance gain compared to all benchmarks.

  \item To enhance system reliability and energy efficiency, we formulate an optimization problem to jointly determine the optimal hovering coordinates of the CMU and the power splitting (PS) factor, achieving the minimum CMU outage probability and maximum EH efficiency. To address this problem, we propose a CMU positioning and PS factor (CPPF) framework as a sub-optimal solution. Although the CPPF method provides good sub-optimal results, it is computationally expensive. To achieve improved performance with lower complexity, we solve the joint CMU position and PS factor optimization via a metaheuristic (JCPPM) algorithm. We compare the proposed algorithms with the {\color{black}RSEL} and heuristic greedy coordinate descent (HGCD) algorithms. The results demonstrate that the proposed algorithms outperform both {\color{black}RSEL} and HGCD benchmarks.

\end{itemize}

\subsection{Organization of the Paper}

The rest of the paper is organized as follows. The proposed hierarchical ad hoc UAV network with non-linear EH and RSMA under the impact of HWI {\color{black}and ICSI} is presented in Section II. We derive the outage probability of the proposed system and the overall outage probability minimization through joint CMU and IoT selection in Section III, while the joint optimization of the UAV position and PS factor is provided in Section IV. Finally, Section V presents and discusses the simulation results, while Section VI offers the conclusion. Finally, Section V presents and discusses the simulation results, while Section VI offers the conclusion. {\color{black}Selected derivations are provided in Appendix A and B. For clarity, the notations used in the rest of the paper are listed in Table~\ref{tab:1}.}






\begin{table}[t]
\centering
\caption{{\color{black}List of Symbols and Descriptions}}
\color{black}
\resizebox{\linewidth}{!}{%
\begin{tabular}{|c|l|}
\hline
\textbf{Symbol} & \textbf{Description} \\
\hline
$\beta$ & Path-loss exponent of the channel\\
\hline
$d_{l}$ & Euclidean distance between the CHU and CMU\\ \hline
$d_{n}$ & Euclidean distance between the CMUs and IoT\\ \hline
$q_{\text{CHU}}$ & 3D position vector of the CHU\\ \hline
$q_{\text{CMU},l}$ & 3D position vector of the $l$-th CMU\\ \hline
$q_{n}$ & 3D position vector of the $n$-th ground IoT device\\ \hline
$H$ & Altitude of the CMUs\\ \hline
$\xi_{1}$, $\xi_{2}$ & Environment-dependent constants \\ \hline
$T$ & Total transmission time \\ \hline
$\rho$ & Power splitting factor \\ \hline
$\eta$ & Energy harvesting efficiency \\ \hline
$P_t$ & Transmit power of the CHU \\ \hline
${g}_{l}$, $\hat{g}_{l}$ & Actual channel and estimated channel coefficients of the CHU–CMU link, respectively\\ \hline
${g}_{n}$, $\hat{g}_{n}$ & Actual channel and estimated channel coefficients of the CMU–IoT link, respectively\\ \hline
$\varepsilon_{l}$, $\varepsilon_{n}$ & Channel estimation error between CHU-CMU and CMU-ground IoT devices, respectively \\ \hline
$m_{l}, m_{n}$ & Spread parameters of the Nakagami-$m$ fading for the $l$-th and $n$-th links\\ \hline
$\Omega_{l}$, $\Omega_{n}$  & Shape parameters of the Nakagami-$m$ fading for the $l$-th and $n$-th links\\ \hline
$\upsilon_{t,l}, \upsilon_{t,n}$ & Distortion noises at the $l$-th and $n$-th transmitters, respectively\\ \hline
$\upsilon_{r,l}, \upsilon_{r,n}$  & Distortion noises at the $l$-th and $n$-th receivers, respectively\\ \hline
$\mu_{l}, \mu_{n}$ & Additive white Gaussian noise at the $l$-th and $n$-th receivers, respectively\\ \hline
$\alpha_{\mathrm{c}}, \alpha_{n}$ & Power allocation coefficients for the common and private signals\\ \hline
${k}_{t,l}, k_{t,n}, {k}_{r,l}, k_{r,n}$ & Levels of impairment at the $l$-th and $n$-th transceivers, respectively\\ \hline
$k_{l}, k_{n}$ & Aggregate level of impairments at the $l$-th and $n$-th transceivers\\ \hline
$K$ & Hardware impairment coefficient \\ \hline
$P_{\text{th}}$ & Saturation threshold\\ \hline
$f_{|h|^{2}}(x)$ & Probability density function\\  \hline
$F_{|h|^{2}}(x)$ & Cumulative distribution function\\ \hline
$\Gamma(m)$ & Gamma function\\ \hline
$\Upsilon(m,x)$ & Lower incomplete Gamma function\\ \hline
$r_{\mathrm{c}}, r_{\mathrm{c}}$ & Target rates of common and private parts, respectively\\ \hline
$\mathcal{X}$ & CMU selection vector\\ \hline
$\mathcal{X}_l$ & Selection variable of the $l$-th CMU\\ \hline
$g_{n}^{\text{fair}}$ & Control factor for the $n$-th ground IoT device, balancing fairness and system performance\\ \hline
$\alpha_{\text{fair}}$ & Weighting coefficient that adjusts the trade-off between fairness and throughput\\ \hline
$\mathcal{K}$ & Required number of CMUs to be selected\\ \hline
$g_{\min}$ & Minimum required channel gain threshold for a CMU to be selected\\ \hline
$\mathcal{M}$ & Large constant used in the Big-M method to deactivate constraints when $\mathcal{X}_l = 0$\\ \hline
$\lambda$ & Penalty coefficient controlling the trade-off between outage minimization and channel quality\\ \hline
$I_{x_{CMU}}, I_{y_{CMU}}, I_{z_{CMU}}, I_\rho$ & Number of discrete levels for the CMU coordinates and the EH factor, respectively\\ \hline
$\mathcal{O}$ & Big-$\mathcal{O}$ notation\\ \hline
$\mathcal{G}$ & Number of generations in the algorithm\\ \hline
$\mathcal{P}$ & Population size in the algorithm\\ \hline
$\mathcal{C}_{\mathrm{eval}}$ & Cost of a single fitness evaluation\\
\hline
\end{tabular}%
}
\label{tab:1}
\end{table}

\section{System and Channel Model}
As illustrated in Fig. 1, we consider a hierarchical ad hoc UAV network consisting of a CHU, $L$ CMUs, indexed by $l=1, 2, ..., L$, and $M$ ground IoT devices, indexed by $n=1, 2, ..., M$. 
The CHU is assumed to serve as the communication source and resource management coordinator for the cluster members, while the cluster members  (i.e., CMUs) act as relay UAVs.
In this scenario, we consider the following assumptions: 1) The CHU is a hovering platform at a fixed location, moving slowly enough to be considered stationary while providing coverage. 2) All nodes are equipped with one (transmit/receive) antenna. 3) {\color{myblue}ICSI} is available at all nodes, and the communication links experience Nakagami-$m$ fading\footnote{{\color{black}Nakagami-$m$ fading is employed due to its ability to provide a flexible description of both LoS and NLoS UAV channel conditions, thereby offering a general model for air-to-ground and air-to-air links~\cite{10836898,10006695}.}}. 4) The UAVs of the cluster members work in the DF protocol and HD mode. 5) HWI is considered at all nodes. 
{\color{myblue}We assume that the CHU is a high-endurance aerial platform with solar-assisted energy support and large onboard energy storage, allowing higher energy availability compared to the CMUs~\cite{10851439}, which supports sustainable network operation. In contrast, the CMUs are supposed to have restricted onboard energy due to their constrained battery capacity. To prolong their operational lifetime, they are equipped with lightweight RF EH modules and storage units, which provide supplementary energy support without significantly increasing payload requirements~\cite{9913422,9707780}.}
In this regard, during each transmission, they harvest a portion of the received signal’s energy using the PS protocol and store it in their batteries, which can then be used to power subsequent transmissions to the ground IoT devices.

Considering the channel scenarios in~\cite{10006695}, the link between the CHU and CMUs in the UAV ad hoc network may establish a direct air-to-air (A2A) connection, which can be considered a LoS link altogether. In contrast, the link between the CMUs and ground IoT devices may establish an air-to-gound (A2G) connection, which can be considered a mixed LoS and non-LoS (NLoS). Thus, the path-loss coefficient of the A2A and A2G links can be formulated, respectively, as~\cite{10836898,10006695}
\begin{equation}
\Upsilon_{l}=  \zeta_{0}  d_{l}^{-\beta}, 
\end{equation}
and
\begin{equation}
\Upsilon_{n}= \left( \zeta_{1} \Lambda_{\text{LoS}} + \zeta_{2} \Lambda_{\text{NLoS}} \right) d_{n}^{-\beta}, 
\end{equation}
where $\zeta_{0}$, $\zeta_{1}$, and $\zeta_{2}$ are the additional path-loss of the channel under LoS and NLoS transmission conditions, $\beta$ is the path-loss exponent of the channel. $d_{l}$ is the Euclidean distance between the CHU and CMUs, which can be written as $\Vert q_{\text{CHU}}-q_{\text{CMU},l}\Vert$, where $q_{\text{CHU}} = (x_{\text{CHU}}, y_{\text{CHU}}, z_{\text{CHU}})$ is the three-dimensional (3D) position vector of the CHU, $q_{\text{CMU},l}= (x_{\text{CMU},l}, y_{\text{CMU},l}, z_{\text{CHU},l})$ is the 3D position vector of the CMUs, and $\Vert.\Vert$ is the Euclidean norm.
$d_{n}$ is the Euclidean distance between the CMUs and ground IoT devices, which can be written as $\Vert q_{\text{CMU},l}-q_{n}\Vert$, where $q_{n}= (x_{n}, y_{n}, z_{n})$ is the 3D position vector of the ground IoT devices. $\Lambda_{\text{LoS}}$ and $\Lambda_{\text{NLoS}}$ are the LoS and NLoS probabilities. Therefore, the probability of LoS channel occurrence can be written as 
\begin{equation}
\Lambda_{\text{LoS}}=\frac{1}{1+ \xi_{1} \exp{\left(-\xi_{2} \left[\arcsin{(\frac{H}{d_{n}})-\xi_{1}} \right]\right)}}, 
\label{eq:3}
\end{equation}
where $H$ is the altitude of the CMUs, $\xi_{1}$ and $\xi_{2}$ are constant values depending on the environment~\cite{10006695}. Accordingly, the probability of NLoS can be given as
\begin{equation}
\Lambda_{\text{NLoS}}=1-\Lambda_{\text{LoS}}. 
\end{equation}

\begin{figure}[!t]
\centering
\includegraphics[width=0.9\columnwidth]{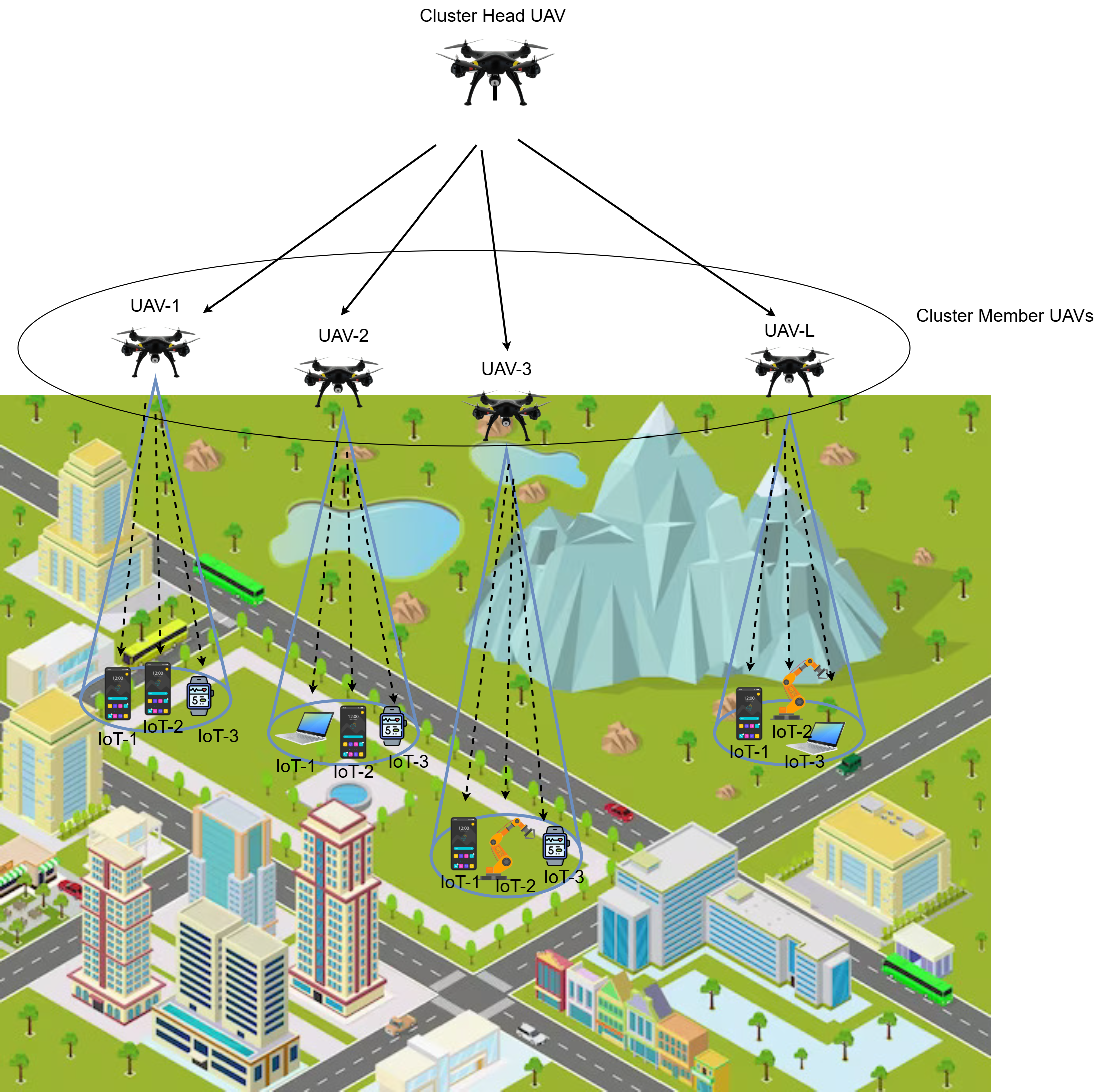}
\caption{Hierarchical ad hoc UAV network with EH system model.}
\label{fig:Fig1}
\end{figure}

\begin{figure}[!t]
\centering
\includegraphics[width=0.9\columnwidth]{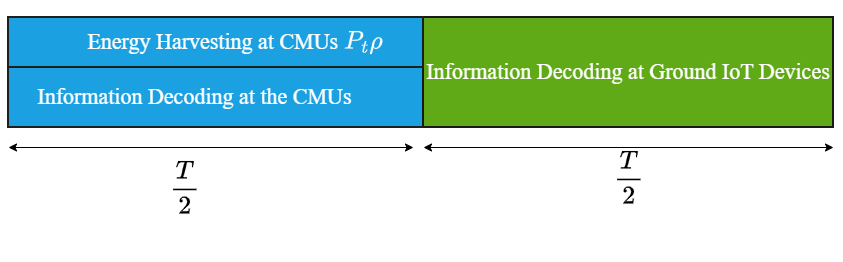}
\caption{Block diagram of the PS protocol for a hierarchical ad hoc UAV network.}
\label{fig:Fig2}
\end{figure}

In the present investigation, we consider a downlink RSMA with PS protocol, in which the CHU serves ground IoT devices through CMUs. In the first and second $\sfrac{T}{2}$ time slots, RSMA is used at CHU and CMUs, where each ground IoT device signal is divided into a common ($W_{\mathrm{c},n)}$ and a private ($W_{\mathrm{p},n}$) parts. The common parts of all ground IoT devices are encoded and combined into the common signal $x_{\mathrm{c}}$, which is a signal that all ground IoT devices must decode. At the same time, private parts are encoded into the private signal $x_{\mathrm{p},n}$. Let $T$ be the total transmission time. The signals are decoded using {\color{black}SIC} during the first and second $\sfrac{T}{2}$ time slots. In the PS protocol, information decoding and EH occur simultaneously. As seen in Fig. 2, in the first $\sfrac{T}{2}$ time slot, the received signal at the CMUs is divided into two parts: one part ($\rho$) is used for EH, and the other part (1-$\rho$) is used for information decoding. The received signal at the CMUs for EH and information decoding can be written by
\begin{equation}
\begin{split}
y_{l}=   (\sqrt{\varsigma P_{t}} X_{S}  + \upsilon_{t,l}) {\color{black}(\hat{g}_{l}+\varepsilon_{l})} + \upsilon_{r,l}+ \mu_{l} ,  \ \varsigma=\{\rho, \iota\},
\end{split}
\end{equation}
where $\iota=(1-\rho)$, $X_{S}=\sum_{n=1}^{M} \sqrt{\alpha_n} x_{\mathrm{p},n} + \sqrt{\alpha_{\mathrm{c}}} x_{\mathrm{c}}$. $\alpha_{\mathrm{c}}$ and $\alpha_{n}$ are the power allocation coefficients for the common signal and the private signal of ground IoT devices, respectively. {\color{black}These coefficients are assumed to be predefined constants\footnote{{\color{black}The power and rate allocation between the common and private streams in the proposed RSMA scheme is considered to be fixed constants as in~\cite{li2025outage,9519666}, i.e., static, for the purpose of analysis of the outage probability. However, dynamic allocation is considered to be beyond the scope of this work and can be addressed in future work.}} satisfying $\alpha_{c}+ \sum_{n=1}^{M} \alpha_{n}=1$ as in~\cite{li2025outage,9519666}.} $g_{l}$ follows instantaneous Nakagami-$m$ fading with $\Omega_{l}$ spread and $m_{l}$ shape parameters, where $\Omega_{l}=\Upsilon_{l}$ and $m_{l}$ are assumed to be greater than 1, as in~\cite{babaei2018ber}. 
{\color{black}Under imperfect CSI, according to the minimum mean-squared error estimation~\cite{9991954,arzykulov2021hardware},
\begin{math}
g_l = \hat{g}_l + \varepsilon_{l},
\end{math}
, where $g_{l}$ is the actual channel between CHU-CMU, $\hat{g}_{l}$ is the estimated channel coefficient between CHU-CMU, and $\varepsilon_{l}$ is the channel estimation error between CHU-CMU, which follows $\mathcal{CN}(0,\sigma_{\varepsilon_{l}}^2)$~\cite{9991954,hoang2020outage}. Assuming that $\hat{g}_{l}$ and $\varepsilon_{l}$ are independent, the estimated channel coefficient is given as \begin{math}
 \hat{g}_l = g_l - \varepsilon_{l}
\end{math}. The average power of the estimated channel between CHU-CMU is therefore given by \begin{math}
\hat{\Upsilon}_l = \Upsilon_l - \sigma_{\varepsilon_l}^2
\end{math}~\cite{9991954,hoang2020outage}}.
$P_{t}$ is the CHU total transmit power, $\rho$ is the PS factor ( $ 0\leq \rho \leq 1$), and $\mu_{l}$ is the additive white Gaussian noise (AWGN), which follows $\ \mu_{l} \sim \mathcal{CN}(0,\ \sigma_{l}^{2})$. $\upsilon_{t,l}$ and $\upsilon_{r,l}$ are distortion noises at the transmitter and receiver, respectively, which occur due to the HWI\footnote{{\color{black}Although HWI is considered in this work, the exploration of mitigation techniques is beyond its scope and will be addressed in future work.}} at transceivers, such as oscillator phase noise, high power amplifier distortion, and in-phase and quadrature-phase imbalance. The distortion noises are defined as \begin{math} \upsilon_{t,l} \sim \mathcal{CN}(0,\ \iota {P}_{t} {k}_{t,{l}}^2 ) \end{math} and \begin{math} \upsilon_{r,l} \sim \mathcal{CN}(0,\ \iota {P}_{t} {k}_{r,{l}}^2 |{g}_{l}|^2) \end{math}, where ${k}_{t,{l}}^2$ and ${k}_{r,{l}}^2$ represent levels of impairment at the transmitter and receiver, respectively~\cite{9991954,10188818}. It was demonstrated in~\cite{9991954,studer2010mimo,10188818} that the impact of the transceiver HWI can be characterized by the aggregate level of impairments and can be written as $k_{l}^2={k}_{t,{l}}^2 + {k}_{r,{l}}^2$.

Considering the non-linear feature of a practical EH circuit, the EH expressions during the EH phase (i.e., first $\sfrac{T}{2}$ time slot) at the CMUs can be given as in~\cite{10188818} by
\begin{equation}
   E_{\text{EH}}=\left \{ \begin{array}{ll}
   \eta \rho P_{t} {\color{black}|\hat{g}_{l}|^2} \frac{T}{2},  & P_{t} {\color{black}|\hat{g}_{l}|^2} \leq P_{\text{th}} ,\\
   \eta \rho P_{\text{th}} \frac{T}{2},  & P_{t} {\color{black}|\hat{g}_{l}|^2} > P_{\text{th}}.
   \end{array} \right.
\end{equation}
where $ 0\leq \eta \leq 1$ is the energy conversion efficiency factor and $P_{\text{th}}$ is the saturation threshold.

The transmission power of the CMUs through the harvested energy is given by
\begin{equation}\label{eq:6}
   P_{\text{EH}}=\frac{ E_{\text{EH}} }{ \sfrac{T}{2} }=\left \{ \begin{array}{ll}
   \eta \rho P_{t} {\color{black}|\hat{g}_{l}|^2} ,  & P_{t} {\color{black}|\hat{g}_{l}|^2} \leq P_{\text{th}} ,\\
   \eta \rho P_{\text{th}} ,  & P_{t} {\color{black}|\hat{g}_{l}|^2} > P_{\text{th}}. 
   \end{array} \right. 
\end{equation}

.


At CMUs, the common signal ($x_{\mathrm{c}}$) is decoded by treating the private signals as noise. Hence, the signal-to-interference-plus-noise ratio (SINR) of the signal $x_{c}$ is given by {\color{black}
\begin{equation}
    \gamma_{l,\mathrm{c}}= \frac{ \iota P_t  \ |{\hat{g}}_{l}|^2   \alpha_{\mathrm{c}}    }{  \iota P_t \ |{\hat{g}}_{l}|^2  \ \sum_{n = 1}^{M}  \  \alpha_{n}  + \iota k_{l}^2   P_{t} |{\hat{g}}_{l}|^2 + \delta_{1} + \sigma_{l}^2}, 
\end{equation}
where \begin{math}
  \delta_{1}=(1+k_{l}^2) \iota P_t \ \sigma_{\varepsilon_l}^2 
\end{math}.
}
After successful decoding of $x_{\mathrm{c}}$, SIC is executed, and the decoded $x_{\mathrm{c}}$ is eliminated from the received signal. The private signal of the ground IoT devices ($x_{\mathrm{p},n}$) is decoded by considering the private signals of all other users as noise. Consequently, the SINR of the signal $x_{\mathrm{p},n}$ can be expressed as {\color{black}
\begin{equation}
    \gamma_{l,n,\mathrm{p}}= \frac{ \iota P_t  \ |{\hat{g}}_{l}|^2   \alpha_{n}    }{  \iota P_t \ |{\hat{g}}_{l}|^2  \ \sum_{ n+1}^{M}  \  \alpha_{n}   + \iota k_{l}^2   P_{t} |{\hat{g}}_{l}|^2 +\delta_{1}  + \sigma_{l}^2}.
\end{equation} 
}

In the remaining time (i.e., the second $\sfrac{T}{2}$ slot), the CMUs re-encode the detected common and private signals similar to the first time slot and forward it to the ground IoT devices using the EH power. The received signal at $n$th ground IoT devices can be written by 
\begin{equation}\label{eq:1}
\begin{split}
y_{n}=   (\sqrt{P_{\text{EH}}} X_{S} + \upsilon_{t,n}) {\color{black}(\hat{g}_{n}+\varepsilon_{n})} + \upsilon_{r,n}+ \mu_{n} ,
\end{split}
\end{equation}
where ${g}_{n}$ follows Nakagami-$m$ fading with spread ${\color{black}\Omega_{n}}=\Upsilon_{n}$ and ${\color{black}m_{n}}$ shape {\color{black}parameters}, which is assumed to be greater than 1, as in \cite{babaei2018ber}. {\color{black}Under imperfect CSI with minimum mean-squared error estimation~\cite{9991954,arzykulov2021hardware}, the channel is modeled as
\begin{math}
g_n = \hat{g}_n + \varepsilon_n
\end{math}, where $g_{n}$ is the actual channel between CMU-ground IoT devices, $\hat{g}_{n}$ is the estimated channel coefficient between CMU-ground IoT devices, and $\varepsilon_{n}$ is the channel estimation error between CMU-ground IoT devices, which follows $\mathcal{CN}(0,\sigma_{\varepsilon_{n}}^2)$~\cite{9991954,arzykulov2021hardware}. Assuming that $\hat{g}_{n}$ and $\varepsilon_{n}$ are independent, the estimated channel coefficient is given as  \begin{math}
 \hat{g}_n = g_n - \varepsilon_{n} \end{math}. The average power of the estimated channel between CMU-ground IoT devices is therefore given by \begin{math}
\hat{\Upsilon}_n = \Upsilon_n - \sigma_{\varepsilon_n}^2
\end{math}~\cite{9991954,hoang2020outage}}.
$\ \mu_{n} \sim \mathcal{CN}(0,\ \sigma_{n}^{2})$ is the AWGN. \begin{math} \upsilon_{t,n} \sim \mathcal{CN}(0,\  {P}_{\text{EH}} {k}_{t,{n}}^2 ) \end{math} and \begin{math} \upsilon_{r,n} \sim \mathcal{CN}(0,\  {P}_{\text{EH}} {k}_{r,{n}}^2 |{g}_{n}|^2) \end{math} are distortion noises at the transceiver, respectively, where $k_{t,n}^{2}$ and $k_{r,n}^{2}$ denote the transceiver impairment levels, respectively. The aggregate HWI is $k_{n}^{2}=k_{t,n}^{2}+k_{r,n}^{2}$~\cite{9991954,studer2010mimo,10188818}.
{\color{black}We consider equal HWI levels as in~\cite{beddiaf2023impact}, i.e., $K=k_{l}=k_{n}$.}

Similar to the first time slot, the common ($x_{\mathrm{c}}$) and private ($x_{\mathrm{p},n}$) signals are detected, respectively, by the $n$th ground IoT device. Accordingly, their SINRs are given by
{\color{black}
\begin{equation}
\begin{split}
    \gamma_{n,\mathrm{c}}=\left \{ \begin{array}{ll} \gamma_{n,\mathrm{c}}^{A} , & P_{t} |{\hat{g}}_{l}|^2 \leq P_{\text{th}} ,\\
    \gamma_{n,\mathrm{c}}^{B} , & P_{t} |{\hat{g}}_{l}|^2 > P_{\text{th}},
    \end{array} \right.
    \end{split}
\end{equation}
and
\begin{equation}
\begin{split}
    \gamma_{n,\mathrm{p}}=\left \{ \begin{array}{ll} \gamma_{n,\mathrm{p}}^{A} , & P_{t} |{\hat{g}}_{l}|^2 \leq P_{\text{th}} ,\\ \gamma_{n,\mathrm{p}}^{B} , & P_{t} |{\hat{g}}_{l}|^2 > P_{\text{th}},
    \end{array} \right.
    \end{split}
\end{equation}

where $\gamma_{n,\mathrm{c}}^{A}=\frac{\eta \rho P_t  \ |{\hat{g}}_{n}|^2 |{\hat{g}}_{l}|^2  \alpha_{\mathrm{c}}    }{ \eta \rho P_t \ |{\hat{g}}_{n}|^2 |{\hat{g}}_{l}|^2 \ \sum_{i = 1}^{M}  \  \alpha_{i}   + k_{n}^2  \eta \rho P_{t} |{\hat{g}}_{n}|^2 |{\hat{g}}_{l}|^2 + \delta_{2}+ \sigma_{n}^2}$
,  $\gamma_{n,\mathrm{c}}^{B}=\frac{\eta \rho P_{th}  \ |{\hat{g}}_{n}|^2   \alpha_{\mathrm{c}}    }{ \eta \rho P_{\text{th}} \ |{\hat{g}}_{n}|^2 \ \sum_{i = 1}^{M}  \  \alpha_{i}   + k_{n}^2  \eta \rho P_{\text{th}} |{\hat{g}}_{n}|^2  + \delta_{3} + \sigma_{n}^2}$,
$\gamma_{n,\mathrm{p}}^{A}=\frac{\eta \rho P_t  \ |{\hat{g}}_{n}|^2 |{\hat{g}}_{l}|^2  \alpha_{n}    }{ \eta \rho P_t \ |{\hat{g}}_{n}|^2 |{\hat{g}}_{l}|^2 \ \sum_{i = n+1}^{M}  \  \alpha_{i}   + k_{n}^2  \eta \rho P_{t} |{\hat{g}}_{n}|^2 |{\hat{g}}_{l}|^2 +\delta_{2} + \sigma_{n}^2}$, $\gamma_{n,\mathrm{p}}^{B}=\frac{\eta \rho P_{\text{th}}  \ |{\hat{g}}_{n}|^2   \alpha_{n}    }{ \eta \rho P_{\text{th}} \ |{\hat{g}}_{n}|^2  \ \sum_{i = n+1}^{M}  \  \alpha_{i}   + k_{n}^2  \eta \rho P_{\text{th}} |{\hat{g}}_{n}|^2 + \delta_{3}  + \sigma_{n}^2}$, \begin{math}
  \delta_{2}=(1+k_{n}^2)  \eta \rho  P_t \ \sigma_{\varepsilon_n}^2 |{\hat{g}}_{l}|^2 
\end{math}, and
\begin{math}
  \delta_{3}=(1+k_{n}^2)\eta \rho P_{\text{th}} \ \sigma_{\varepsilon_n}^2 
\end{math}.
}









\section{Outage Probability analysis}
This section derives the outage probability of the hierarchical ad hoc UAV network with non-linear EH by calculating the analytical expressions for each ground IoT device within one CMU. We take into account the effect of non-ideal conditions (i.e., HWI {\color{black}and ICSI}) in our derivation. Furthermore, we analyze the approximation behavior of the outage probability at high transmit power regimes for ground IoT devices to gain a comprehensive understanding of the proposed system. Then, we obtain the outage probability of the system at a single CMU and the overall outage probability of a hierarchical ad hoc UAV network (i.e., outage probability at $L$ CMUs).

The probability density function (PDF) and cumulative distribution function (CDF) of $h$ can be, respectively, given as
\begin{equation}
f_{|h|^{2}}(x)= \left( \frac{m}{\Omega}\right)^{m} \frac{x^{m-1}}{\Gamma(m)} e^{-\frac{m}{\Omega}x}
\end{equation}
and
\begin{equation}
\begin{split}
F_{|h|^{2}}(x)= \frac{\Upsilon(m,\frac{m}{\Omega}x)}{\Gamma(m)},
\end{split}
\end{equation}
where \begin{math}\Gamma(m) =\int_{0}^{\infty} x^{m-1} e^{-x} dx\end{math} and \begin{math}\Upsilon(m,x) =\int_{0}^{x} x^{m-1} e^{-x} dx\end{math} represent the Gamma function and lower incomplete Gamma function, respectively.

\subsection{Outage Probability of RSMA devices}
In the first and second time slots, the outage will never occur if CMUs and ground IoT devices are able to successfully decode both the private message $x_{p,n}$, and the common message $x_{c}$. In particular, the end-to-end (E2E) outage probability of the $n$th ground IoT device is expressed by
\begin{equation}
    P_{n}(\text{out}) = 1-\left(1- P_{l, n}^{I} (\text{out}) \right) \left( 1- P_{n}^{II} (\text{out}) \right),
\end{equation}
where $P_{l, n}^{I}(\text{out})$ and  $P_{n}^{II}(\text{out})$ denote the outage probabilities at the CMU and ground IoT device corresponding to the first and second $\sfrac{T}{2}$ slots, respectively. The outage probability at the CMU {\color{black}under the impact of HWI and ICSI} is given by
\begin{equation}
    P_{l, n}^{I}(\text{out})= \frac{\Upsilon(m_{l},\frac{m_{l}}{\Omega_{l}}\hat{\gamma}_{l,n})}{\Gamma(m_{l})},
 \end{equation} 
{\color{black}where
$\hat{\gamma}_{l,n}=\max\{ \mho_{l,\mathrm{c}}, \mho_{l,n,\mathrm{p}}\}$,
\begin{math}
     \mho_{l,\mathrm{c}}= \frac{ (\delta_{1}+\sigma_{l}^2) \ \varphi_{l,\mathrm{c}} }{( \alpha_{\mathrm{c}} - \sum_{i = 1}^{M}  \  \alpha_{i}  \ \varphi_{l,\mathrm{c}} +  k_{l}^2 \ \varphi_{l,\mathrm{c}} ) \ \iota \ P_t}    
\end{math}, and
 \begin{math}
    \mho_{l,n,\mathrm{p}}= \frac{ (\sigma_{l}^2 +\delta_{1}) \ \varphi_{l,n}}{(\alpha_{n}- \sum_{i = n+1}^{M}  \  \alpha_{i}  \ \varphi_{l,n}  +  k_{l}^2 \ \varphi_{l,n}) \ \iota P_t }   
 \end{math}.}
$\varphi_{l,\mathrm{c}}=2^{2r_{\mathrm{c}}}-1$ and $\varphi_{l,n}=2^{2r_{\mathrm{p}}}-1$, where $r_{\mathrm{c}}$ and $r_{\mathrm{c}}$ denote the target rates of $x_{\mathrm{c}}$ and $x_{\mathrm{p},i}$, respectively.


\textit{\textbf{Theorem 1}}: The outage probability at the $n^{th}$ ground IoT device occurs when the common ($x_{\mathrm{c}}$) or private ($x_{\mathrm{p},n}$) messages are not successfully decoded. {\color{black}Hence, by taking into account the impact of HWI and ICSI, the outage probability during the second $\sfrac{T}{2}$ time slot can be expressed as}
\begin{equation}
\begin{split}
& P_{n}^{II} (\text{out})=  \\& 1- \prod_{n}^{M}  \ \left(1-   \aleph_{1}  -\aleph_{2} (1 -  \aleph_{3})  \right)   
  \times 
 \left( 1-  \aleph_{1} -\aleph_{5}(1 - \aleph_{6} )\right) ,
\end{split}
\end{equation}
where  
\begin{math}
\aleph_{1}=\int_{0}^{\frac{P_{\text{th}}}{P_{t}}} \frac{\mathfrak{L}_{1}}{\Gamma(m_{l})} 
    \cdot\mathfrak{L}_{2} \cdot e^{-\frac{m_{n} }{\Omega_{n}}x} dx,
\end{math}
\begin{math}
\aleph_{2}=\frac{\Upsilon(m_{n},\frac{m_{n}}{\Omega_{n}}\mho_{n,\mathrm{c}}^{B})}{\Gamma(m_{n})},
\end{math}
\begin{math}
\aleph_{3}=\frac{\Upsilon(m_{l},\frac{m_{l}}{\Omega_{l}}\chi)}{\Gamma(m_{l})},
\end{math} 
\begin{math}
\aleph_{4}=\int_{0}^{\frac{P_{\text{th}}}{P_{t}}} \frac{\mathfrak{L}_{3}}{\Gamma(m_{l})} 
    \cdot
    \mathfrak{L}_{2} \cdot e^{-\frac{m_{n} }{\Omega_{n}}x} dx,
\end{math}
\begin{math}
\aleph_{5}=\frac{\Upsilon(m_{n},\frac{m_{n}}{\Omega_{n}}\mho_{n,\mathrm{p}}^{B})}{\Gamma(m_{n})},
\end{math}
\begin{math}
\aleph_{6}=\frac{\Upsilon(m_{l},\frac{m_{l}}{\Omega_{l}}\chi)}{\Gamma(m_{l})},
 \end{math}
 \begin{math}
    \chi=\frac{P_{\text{th}}}{P_{t}},
\end{math}
 \begin{math}\mathfrak{L}_{1}=\Upsilon\left(m_{l}, \frac{m_{l} \mho_{n,\mathrm{c}}^{A}}{\Omega_{l} x}\right),
 \end{math}
\begin{math}\mathfrak{L}_{2}=\frac{m_{n}^{m_{n}} x^{m_{n} - 1}}{\Gamma(m_{n}) \Omega_{n}^{m_{n}}} ,
\end{math} and
\begin{math}\mathfrak{L}_{3}=\Upsilon\left(m_{l}, \frac{{\color{black}m_{l}} \mho_{n,\mathrm{p}}^{A} }{\Omega_{l} x}\right),
\end{math} in which

{\color{black}
 \begin{align}
          \mho_{n,\mathrm{c}}^{A}= \frac{(\sigma_{n}^2 + \mathfrak{z}_1 x) \varphi_{n,c}}{( \alpha_{\mathrm{c}} - \varphi_{n,\mathrm{c}}  \ \sum_{i = 1}^{M}  \  \alpha_{i}   + k_{n}^2  \varphi_{n,\mathrm{c}} ) \eta \rho P_t }  , 
\end{align}
\begin{align}
     \mho_{n,\mathrm{c}}^{B}= \frac{ (\sigma_{n}^2 + \mathfrak{z}_2)\varphi_{n,\mathrm{c}}}{(\alpha_{c} - \sum_{i = 1}^{M}  \  \alpha_{i}  \varphi_{n,\mathrm{c}} + k_{n}^2  \varphi_{n,\mathrm{c}} ) \eta \rho P_{\text{th}} } ,
\end{align}
\begin{align}
     \mho_{n,\mathrm{p}}^{A}= \frac{ (\sigma_{n}^2 + \mathfrak{z}_1 x)  \varphi_{n}}{(\alpha_{n} - \varphi_{n} \sum_{i = n+1}^{M}  \  \alpha_{i}   + k_{n}^2 \varphi_{n}) \eta \rho P_{t} },
\end{align}
\begin{align}
     \mho_{n,\mathrm{p}}^{B}= \frac{ (\sigma_{n}^2 + \mathfrak{z}_2) \varphi_{n}}{(\alpha_{n} - \varphi_{n} \ \sum_{i = n+1}^{M}  \  \alpha_{i}   + k_{n}^2 \varphi_{n}) \eta \rho P_{\text{th}} },
\end{align}}
where
{\color{black}
\begin{math}
    \mathfrak{z}_1=(1+k_{n}^2)  \eta \rho  P_t \ \sigma_{\varepsilon_n}^2
\end{math}, 
and \begin{math}
    \mathfrak{z}_2=(1+k_{n}^2)\eta \rho P_{\text{th}} \ \sigma_{\varepsilon_n}^2
\end{math}.}

\begin{IEEEproof} Please see {\color{black}Appendix A}.\end{IEEEproof}

Hence, to find the E2E outage probability of the $n$th ground IoT device, we substitute (16) and (17) into (15).

\textit{\textbf{Corollary 1}}: To gain further insights into the performance of the system, the approximate outage probability is derived for the case when $P_{t} \rightarrow  \infty$. To obtain the approximation of the E2E outage probability for the $n$th ground IoT device, we determine the approximation of the outage probability at the first and second time slots (i.e., approximation of (16) and (17)) and then substitute them into (15).
The {\color{black}approximation of the outage probability} at the first $\sfrac{T}{2}$ time slot (i.e., at the CMU) {\color{black}under the impact of HWI and ICSI} can be expressed~as
\begin{equation}
    P_{l, n}^{I, \infty}(\text{out})\approx \frac{ \left( \frac{m_{l}}{\Omega_{l}} \hat{\gamma}_{l,n} \right)^{m_{l}} }{\Gamma(m_{l}) m_{l}} ,
 \end{equation} 
and the {\color{black}approximation of the outage probability} at the second $\sfrac{T}{2}$ time slot (i.e., at the ground IoT devices) {\color{black}under the impact of HWI and ICSI} can be expressed as 
\begin{equation}\label{eq:17}
\begin{split}
& P_{n}^{II, \infty} (\text{out}) \approx \\& 1- \prod_{n}^{M}  \  \left( 1-   \mathfrak{C}_{1} - \mathfrak{C}_{2} (1  - \mathfrak{C}_{3})  \right) 
 \times 
 \left( 1-  \mathfrak{C}_{4}    - \mathfrak{C}_{5} (1- \mathfrak{C}_{6}   \right),
\end{split}
\end{equation}
{\color{black}where 
\begin{math}
 \mathfrak{C}_{1}=\mathfrak{F}_1  \left(\frac{\Omega_n}{m_n}\right)^{m_n - m_l}
    \mathfrak{T}_1,
\end{math} 
\begin{math}
  \mathfrak{C}_{2}= \frac{ (\frac{m_{n}}{\Omega_{n}}\mho_{n,\mathrm{c}}^{B})^{m_{n}} }{\Gamma(m_{n}) m_{n}} ,
\end{math}
\begin{math}
   \mathfrak{C}_{3}=\frac{(\frac{m_{l}}{\Omega_{l}}\chi)^{m_{l}}}{\Gamma(m_{l}) m_{l}},
 \end{math}
\begin{math}
\mathfrak{C}_{4}=\mathfrak{F}_2  \left(\frac{\Omega_n}{m_n}\right)^{m_n - m_l}
    \mathfrak{T}_2,
\end{math}
\begin{math}
 \mathfrak{C}_{5}=\frac{ (\frac{m_{n}}{\Omega_{n}}\mho_{n,\mathrm{p}}^{B})^{m_{n}} }{\Gamma(m_{n}) m_{n}} ,
\end{math}
\begin{math}
\mathfrak{C}_{6}=\frac{(\frac{m_{l}}{\Omega_{l}}\chi).^{m_{l}}}{\Gamma(m_{l}) m_{l}},
\end{math}
\begin{math}
    \mathfrak{F}_1=\frac{ m_{n}^{m_{n}} \left(   \mho_{n,\mathrm{c}}^{A_1}   \right)^{m_{l}}  }{m_{l} \Gamma(m_{l}) \Gamma(m_{n}) \Omega_{n}^{m_{n}}},
\end{math}
\begin{math}
    \mathfrak{F}_2=\frac{ m_{n}^{m_{n}} \left(   \mho_{n,\mathrm{p}}^{A_1}   \right)^{m_{l}}  }{m_{l} \Gamma(m_{l}) \Gamma(m_{n}) \Omega_{n}^{m_{n}}},
\end{math}
\begin{math}
   \mathfrak{T}_1= \left( \mathfrak{V_1} \mathfrak{U}_1 + \mathfrak{V_2} \mathfrak{U}_2 ,
      \right)
\end{math}
\begin{math}
    \mathfrak{T}_2=\left( \mathfrak{V_1} \mathfrak{U}_3 + \mathfrak{V_2} \mathfrak{U}_4 ,
      \right)
\end{math}
\begin{math}
    \mathfrak{U}_1=\frac{  \left(   \mho_{n,\mathrm{c}}^{A_1}   \right)^{m_{l}}  }{m_{l} },
\end{math}
\begin{math}
    \mathfrak{U}_2=\frac{ \mho_{n,\mathrm{c}}^{A_2} }{ \mho_{n,\mathrm{c}}^{A_1} } 
    \frac{\Omega_n}{m_n},
\end{math}
\begin{math}
    \mathfrak{U}_3=\frac{  \left(   \mho_{n,\mathrm{p}}^{A_1}   \right)^{m_{l}}  }{m_{l} } ,
\end{math}
\begin{math}
    \mathfrak{U}_4=\frac{ \mho_{n,\mathrm{p}}^{A_2} }{ \mho_{n,\mathrm{c}}^{A_1} } 
    \frac{\Omega_n}{m_n},
\end{math}
\begin{math}\mathfrak{V_1}= 
    \Upsilon\left(m_n - m_l, \frac{m_n P_{\text{th}}}{\Omega_n P_{\text{t}}} \right),
\end{math}
\begin{math}
\mathfrak{V_2}= \Upsilon\left(m_n - m_l +1, \frac{m_n P_{\text{th}}}{\Omega_n P_{\text{t}}} \right),
\end{math} in which

\begin{align}
     \mho_{n,\mathrm{c}}^{A_1}= \frac{ m_{l} \sigma_{n}^2 \varphi_{n,c}}{ \Omega_{l}( \alpha_{\mathrm{c}} - \varphi_{n,\mathrm{c}}  \ \sum_{i = 1}^{M}  \  \alpha_{i}   + k_{n}^2  \varphi_{n,\mathrm{c}} ) \eta \rho P_t } ,
\end{align}
\begin{align}
     \mho_{n,\mathrm{c}}^{A_2}=  \frac{ m_{l} \mathfrak{z}_1  \varphi_{n,c}}{ \Omega_{l}( \alpha_{\mathrm{c}} - \varphi_{n,\mathrm{c}}  \ \sum_{i = 1}^{M}  \  \alpha_{i}   + k_{n}^2  \varphi_{n,\mathrm{c}} ) \eta \rho P_t },
\end{align}
\begin{align}
     \mho_{n,\mathrm{p}}^{A_1}= \frac{ m_{l} \sigma_{n}^2   \varphi_{n}}{ \Omega_{l} (\alpha_{n} - \varphi_{n} \sum_{i = n+1}^{M}  \  \alpha_{i}   + k_{n}^2 \varphi_{n}) \eta \rho P_{t} },
\end{align}
\begin{align}
     \mho_{n,\mathrm{p}}^{A_2}= \frac{ m_{l}  \mathfrak{z}_1   \varphi_{n}}{ \Omega_{l}(\alpha_{n} - \varphi_{n} \sum_{i = n+1}^{M}  \  \alpha_{i}   + k_{n}^2 \varphi_{n}) \eta \rho P_{t} }.
\end{align}
}

\begin{IEEEproof} The proof of (23) is provided in {\color{black}Appendix B}.\end{IEEEproof}

Hence, to find the approximation of the E2E outage probability of the $n^{th}$ ground IoT device, we substitute (22) and (23) into (15).

\subsection{CMU Outage Probability}
A single CMU is considered in outage if the transmission to at least one associated ground IoT device fails, i.e., if the device cannot correctly decode its intended signal. Thus, the CMU outage probability is expressed by
\begin{equation}
    P_{\text{CMU},l} (\text{out})= 1- \prod_{n}^{M} \left(1- P_{ n} (\text{out})\right) ,
\end{equation}
where $P_{ n} (\text{out})$ is the outage probability of the $n$th ground IoT device, which is given in (15).

\subsection{Overall Outage Probability}
The overall outage probability denotes the probability that at least one CMU system fails to serve its ground IoT devices. Hence, the overall outage probability of $l$ CMUs can be formulated by
\begin{equation}
    P_{\text{sys}} (\text{out})= 1- \prod_{l}^{L} \left(1- P_{\text{CMU},l} (\text{out}) \right) .
\end{equation}

\textit{Remark:} It is observed from (29) that increasing the number of CMUs, $L$, leads to a degradation in outage performance. This is an expected outcome, as the overall system reliability is limited by its weakest link. As the number of CMUs increases, the probability that at least one CMU experiences an outage and fails to serve its ground IoT devices also increases.
Even if each CMU has a low chance of individual outage, aggregating the risk across many clusters leads to an accumulation of total outages. The service area or ground IoT devices population can be expanded by adding additional CMUs, but the system is more vulnerable to individual connection failures. This impact represents a trade-off between coverage and reliability\footnote{If the area of CHU is fixed, increasing the number of CMUs to increase coverage and serve a wider area means increasing the footprint of CHU, which leads to a decrease in the quality of service~\cite{kurt2021vision}.}.



\subsubsection{Joint User Pairing and Soft-Constrained CMU Selection}

To address the trade-off between coverage and reliability, we formulate a problem of selecting $\mathcal{K}$ CMUs out of $L$ candidates such that the overall system outage probability is minimized. 
Given that each CMU accommodates multiple ground IoT devices, we initially conduct ground IoT device pairing within each CMU by assessing all potential pairs of candidate ground IoT devices $(\mathfrak{i}, \mathfrak{j})$, where $\mathfrak{i} \neq \mathfrak{j}$ and identifying the pair that results in the minimum CMU outage probability. For each pair, ground IoT devices are dynamically assigned as near and far based on their respective channel conditions, i.e., $g_{\mathfrak{i}} \leq g_{\mathfrak{j}}$. The pair $(\mathfrak{i}^*, \mathfrak{j}^*)$ that minimizes the CMU outage probability is selected according to
\begin{equation}
    (\mathfrak{i}^*, \mathfrak{j}^*) = \arg \min_{\mathfrak{i} \ne \mathfrak{j}}  P_{\text{CMU},l} (\mathfrak{i}, \mathfrak{j}),
\end{equation}
where $P_{\mathrm{CMU},l}  (\mathfrak{i}, \mathfrak{j})$ denotes the CMU outage probability for user pair $(\mathfrak{i}, \mathfrak{j})$ in the $l$-th CMU, which is defined in (20). 
The outage corresponding to the optimal ground IoT device pair in each CMU is stored, along with the associated ground IoT device indices. These values are used as a representative performance metric for the subsequent system-level selection process. 
To further enhance reliability, we incorporate a soft constraint based on the minimum ground IoT device channel gain within each CMU, penalizing CMUs whose minimum gain falls below a predefined threshold $g_{\min}$. 
Let $\mathcal{X} = [\mathcal{X}_1, \mathcal{X}_2, \ldots, \mathcal{X}_L]^{\mathsf{T}}$ denote the CMU selection vector, where each selection variable \(\mathcal{X}_l\) is a relaxed version of the original binary decision variable, such that $\mathcal{X}_l \in [0,1]$ and
\begin{equation}
 \mathcal{X}_l =
\begin{cases}
1, & \text{if the $l$-th CMU is selected},\\[6pt]
0, & \text{otherwise}.
\end{cases}   
\end{equation}

Hence, the problem can be formulated as 
\begin{subequations}
\begin{align}
(\text{P1}): \quad & \operatorname*{min}_{\mathcal{X}} \quad P_{\text{sys}} (\text{out}) \\
\text{s.t.:} \quad 
& \sum_{l=1}^{L} \mathcal{X}_{l} = \mathcal{K}, \\
& 0 \leq \mathcal{X}_{l} \leq 1, \quad \forall l, \\
& |g_{n}^{\text{fair}}|^2 \geq |g_{\text{min}}|^2 - \mathcal{M}(1 - \mathcal{X}_{l}), \quad \forall n,
\end{align}
\end{subequations}
where $g_{n}^{\text{fair}}=\alpha_{\text{fair}} \min(g_{n}^{(\mathfrak{i})}, g_{n}^{(\mathfrak{j})}) + (1 - \alpha_{\text{fair}}) \max(g_{n}^{(\mathfrak{i})}, g_{n}^{(\mathfrak{j})})$, $g_{n}^{\text{fair}}$ is the control factor\footnote{In order to strike a good balance between system performance and fairness during ground IoT device pairing, $g_{n}^{\text{fair}}$ is the weighting of the minimum and maximum channel gains for each ground IoT device pair inside each CMU. This way, weaker ground IoT devices won't be ignored, and the system will be very efficient overall. The balance between prioritizing fairness and maximizing throughput is controlled by adjusting $\alpha_{\text{fair}}$.} that balances the trade-off between fairness and performance, $P_{\text{sys}}(\text{out})$ is the overall system outage probability, which is the objective function to be minimized, $\mathcal{X}_l$ is the relaxed selection variable for the $l$-th CMU, in which $\mathcal{X}_l \in [0,1]$. $L$ is the total number of CMUs and $\mathcal{K}$ is the required number of CMUs to be selected. $g_{n}^{(\mathfrak{i})}$ and $g_{n}^{(\mathfrak{j})}$ are channels of the selected ground IoT devices in $l$th CMU, while $g_{\min}$ is the minimum required channel gain threshold for a CMU to be selected. $\mathcal{M}$ is a large constant used in the Big-M method~\cite{wolsey1999integer,trespalacios2015improved} to deactivate the constraint when $\mathcal{X}_l = 0$.




{\color{black}P1 is inherently difficult to solve due to its mixed discrete–continuous nature and non-linear objective. The CMU selection variables $\mathcal{X}_l$ are binary, and together with continuous optimization variables, this makes (P1) a mixed-integer non-linear programming (MINLP) problem. Moreover, the objective function defined in (32) involves a product of nonlinear outage probabilities, which depend on exponential and incomplete Gamma functions from the Nakagami-m model. Consequently, the objective function is non-convex with respect to the optimization variables. Additionally, the Big-$M$ constraint introduces further nonlinearity and couples selection variables with channel gains. Due to the combinatorial nature of the binary selection variables and the nonlinear product structure of the outage probability expression, (P1) is non-convex and non-deterministic polynomial time (NP-hard) in general, making direct optimization challenging~\cite[Sec.~1]{pardo2020statistical}.}
To make the problem tractable and compatible with convex optimization solvers such as CVX, the binary constraint on $\mathcal{X}_{l}$ is relaxed to a continuous domain, i.e., $\mathcal{X}_{l} \in [0,1]$. After this relaxation, (P1) becomes a continuous nonlinear programming (NLP) problem.
Moreover, the multiplication formula makes the optimization method difficult, especially in cases where $\mathcal{X}_{l}$ is a continuous relaxation. Therefore, to make the problem solvable using convex optimization approaches, we reformulate it using a logarithmic transformation. Hence, after substituting (29) into (P1) and some mathematical manipulations, the problem can be reformulated as 
\begin{subequations}
\begin{align}
(\text{P2}): \quad & \operatorname*{min}_{\mathcal{X}} \quad - \sum_{l=1}^{L} \mathcal{X}_l \log\left(1 -  P_{\mathrm{CMU},l}  (\text{out}) \right) \\
\text{s.t.:} \quad 
& \text{(32b), (32c), (32d)}.
\end{align}
\end{subequations}

However, the Big-M constraint~(32d) introduces numerical challenges and may render the problem infeasible when no CMUs strictly satisfy the minimum channel gain threshold. To address these issues, we propose replacing the hard constraint with a soft constraint formulation by incorporating a soft penalty into the objective function~\cite[Sec.~17.1]{nocedal2006numerical}. The reformulated problem with this penalty term ensures flexibility and numerical stability, and is expressed as
\begin{subequations}
\begin{align}
(\text{P3}): \quad & \operatorname*{min}_{\mathcal{X}} \quad - \sum_{l=1}^{L} \mathcal{X}_l \log\left(1 -  P_{\mathrm{CMU},l}  (\text{out}) \right) + \Psi \\
\text{s.t.:} \quad 
& \text{(32b), (32c)},
\end{align}
\end{subequations}
where \begin{math}\Psi=\lambda \sum_{l=1}^L \mathcal{X}_l \max\left(0, |g_{n}^{\text{fair}}|^2 - |g_{min}|^2 \right)\end{math}, \begin{math} \lambda > 0 \end{math} is a penalty coefficient controlling the trade-off between outage minimization and channel quality. The penalty term softly discourages the selection of CMUs with gains below the threshold without strictly prohibiting them, improving numerical stability and feasibility. The steps are summarized in~Algorithm 1.

\begin{algorithm}[]
\caption{OPM-JCIS Algorithm}
\begin{algorithmic}[1]

\State \textbf{Initialize parameters:} $L,K,\mathcal{M},g_{\min}$

\For{$\mathsf{i}=1$ to $P_t$}

    \State $P_s \leftarrow P_t(\mathsf{i})$
    
    \State Initialize $P_{sys,l}(\text{out}) \leftarrow \mathbf{0}$

    \For{$l=1$ to $L$}

        \State $P_{sys,best}(\text{out}) \leftarrow 1$

        \For{$\mathfrak{i}=1$ to $M$}

            \For{$\mathfrak{j}=1$ to $M$}

                \If{$\mathfrak{i}=\mathfrak{j}$}
                    \State \textbf{continue}
                \EndIf

                \State Order channel gains $g_n$ such that $g_{\mathfrak{i}} \leq g_{\mathfrak{j}}$

                \State Compute: $P_n(\text{out})$ and $P_{\text{CMU},l}(\text{out})$

                \If{$P_{\text{CMU},l}(\text{out}) < P_{\text{sys,best}}(\text{out})$}

                    \State Update $P_{\text{sys,best}}(\text{out}) \leftarrow P_{\text{CMU},l}(\text{out})$

                    \State Store selected users indices $[\mathfrak{i},\mathfrak{j}]$

                \EndIf

            \EndFor

        \EndFor

        \State $P_{\text{sys}}(\text{out})(l) \leftarrow P_{\text{sys,best}}(\text{out})$

    \EndFor

    \State Store $P_{\text{sys}}(\text{out})(l)$ for current $P_s$

    \State Create the optimization using (34)

    \State Select top $\mathcal{K}$ CMUs based on $\mathcal{X}$

    \State Calculate $P_{\text{sys}}(\text{out})$ for selected CMUs using (29)

\EndFor

\end{algorithmic}
\end{algorithm}

{\color{black}
\textbf{Computational Complexity Analysis:}
The proposed OPM-JCIS algorithm has a computational complexity of approximately $\mathcal{O}(LM^{2}+L)$, where $L$ represents the number of candidate CMUs and $M$ denotes the number of IoT devices associated with each CMU. The quadratic term $M^{2}$ arises because all possible IoT device pairs must be evaluated within each CMU in order to determine the pair that minimizes the outage probability. Since this procedure is repeated for all $L$ CMUs, the resulting complexity becomes $\mathcal{O}(LM^{2})$. In addition, the final CMU selection step requires evaluating the objective function once for each CMU, which introduces an additional linear complexity of $\mathcal{O}(L)$. Therefore, the overall computational complexity of the OPM-JCIS algorithm can be expressed as
\begin{math}
\mathcal{O}(LM^{2}+L) \approx \mathcal{O}(LM^{2}),
\end{math}
where the quadratic term dominates the linear term according to standard Big-$\mathcal{O}$ asymptotic analysis, which neglects lower-order terms and constant factors~\cite{thomas2009introduction}.
}

\section{Joint UAV Position and PS Factor Optimization }
The CMU position is a key factor in improving the performance of ground IoT devices and EH efficiency. The PS factor $\rho$ is another important factor in determining the amount of harvested energy and the quality of the received signal. In this section, we aim to jointly optimize the 3D position of a single CMU and PS factor $\rho$ to minimize the CMU outage probability. In this scenario, we assume that the CHU and the ground IoT devices have fixed locations, while the CMU is allowed to hover and move within a 3D space ${\color{black}(x_{CMU}, y_{CMU}, z_{CMU})}$. 
Hence, the optimization problem can be formulated as
\begin{subequations}
\begin{align}
(\text{P4}): \quad & \operatorname*{min}_{\mathcal{Y}} \quad P_{\text{CMU},l} (\text{out}) \\
\text{s.t.:} \quad 
& {\color{black}x_{\min} \leq x_{CMU} \leq x_{\max}, \quad, }\\
&{\color{black} y_{\min} \leq y_{CMU} \leq y_{\max}, \quad, }\\
& {\color{black}z_{\min} \leq z_{CMU} \leq z_{\max}, \quad, }\\
& \rho_{\min} \leq \rho \leq \rho_{\max},
\end{align}
\end{subequations}
where $P_{\text{CMU},l} (\text{out})$ represents the CMU outage probability, which is based on the PS factor and CMU’s 3D position, $\mathcal{Y} = {\color{black}( x_{CMU}, y_{CMU}, z_{CMU}, \rho )}$ is the vector of optimization variables, in which {\color{black}$x_{CMU}$, $y_{CMU}$, $z_{CMU}$}, and $\rho$ represent the 3D coordinates of the CMU position and PS factor. $x_{\min}$, $x_{\max}$, $y_{\min}$, $y_{\max}$, $z_{\min}$, $z_{\max}$, $\rho_{\min}$, and $\rho_{\max}$ represent the minimum and maximum bounds of the CMU’s 3D coordinates ${\color{black}(x_{CMU}, y_{CMU}, z_{CMU})}$ and $\rho$, respectively.

\begin{algorithm}[]
\caption{CPPF Algorithm}
\begin{algorithmic}[1]

\State \textbf{Initialize:} 
$x_{\text{CMU}} \in [x_{\min}, x_{\max}], \quad x_{\text{CMU}} \ne x_{\text{CHU}},
y_{\text{CMU}} \in [y_{\min}, y_{\max}], \quad y_{\text{CM}} \ne y_{\text{CHU}},
z_{\text{CMU}} \in [z_{\min}, z_{\max}], \quad z_{\text{CMU}} < z_{\text{CHU}},
\rho \in [\rho_{\min}, \rho_{\max}], \quad \rho \ne 0$

\State Set $P_{\mathrm{CMU},l}^{\text{min}}(\text{out}) \leftarrow \infty$

\For{$\mathsf{i}=1$ to $\mathrm{length}(x_{\mathrm{CMU}})$}

    \State $x_{\mathrm{CMU},\mathsf{i}} \leftarrow x_{\mathrm{CMU}}(\mathsf{i})$

    \If{$x_{\mathrm{CMU},\mathsf{i}} = x_{\mathrm{CHU}}$}
        \State \textbf{continue}
    \EndIf

    \For{$\mathsf{j}=1$ to $\mathrm{length}(y_{\mathrm{CMU}})$}

        \State $y_{\mathrm{CMU},\mathsf{j}} \leftarrow y_{\mathrm{CMU}}(\mathsf{j})$

        \If{$y_{\mathrm{CMU}} = y_{\mathrm{CHU}}$}
            \State \textbf{continue}
        \EndIf

        \For{$\mathsf{k}=1$ to $\mathrm{length}(z_{\mathrm{CMU}})$}

            \State $z_{\mathrm{CMU},\mathsf{k}} \leftarrow z_{\mathrm{CMU}}(\mathsf{k})$

            \If{$z_{\mathrm{CMU}} \geq z_{\mathrm{CHU}}$}
                \State \textbf{continue}
            \EndIf

            \For{$\mathsf{f}=1$ to $\mathrm{length}(\rho)$}

                \State $\rho_{\mathsf{f}} = \rho(\mathsf{f})$

                \State Compute $P_{\text{CMU},l}(\text{out})$ using (24)

                \If{$P_{\text{CMU},l}(\text{out}) < P_{\text{CMU},l}^{\text{min}}(\text{out})$}

                    \State Update $P_{\text{CMU},l}^{\text{min}}(\text{out}) \leftarrow P_{\text{CMU},l}(\text{out})$

                    \State Store current best values: {\color{black}$x^*_{\mathrm{CMU}} \leftarrow x_{\mathrm{CMU}}$, $y^*_{\mathrm{CMU}} \leftarrow y_{\mathrm{CMU}}$, $z^*_{\mathrm{CMU}} \leftarrow z_{\mathrm{CMU}}$, $\rho^* \leftarrow \rho$}

                \EndIf

            \EndFor

        \EndFor

    \EndFor

\EndFor

\State \textbf{Return:} Optimal
$(x^*_{\mathrm{CMU}}, y^*_{\mathrm{CMU}}, z^*_{\mathrm{CMU}}), \rho^*$, and $P_{\text{CMU},l}^{\text{min}}(\text{out})$

\end{algorithmic}
\end{algorithm}

\begin{algorithm}[]
\caption{JCPPM Algorithm}
\begin{algorithmic}[1]

\State \textbf{Initialize:} 
$x_{\mathrm{CMU}} \in [x_{\min}, x_{\max}], \quad x_{\mathrm{CMU}} \ne x_{\mathrm{CHU}},
y_{\mathrm{CMU}} \in [y_{\min}, y_{\max}], \quad y_{\mathrm{CMU}} \ne y_{\mathrm{CHU}},
z_{\mathrm{CMU}} \in [z_{\min}, z_{\max}], \quad z_{\mathrm{CMU}} < z_{\mathrm{CHU}},
\rho \in [\rho_{\min}, \rho_{\max}], \quad \rho \ne 0$

\State Set evolutionary control parameters: $\mathcal{G}$ and $\mathcal{P}$

\For{$\mathsf{i} = 1:\mathrm{length}(P_t)$ ; \quad $P_t = P_t(\mathsf{i})$}

    \State Define objective function:
    \begin{align*}
    f(x_{\mathrm{CMU}}, y_{\mathrm{CMU}}, z_{\mathrm{CMU}}, \rho)
    = & \\ P_{\mathrm{CMU},l}(P_t, x_{\mathrm{CMU}}, y_{\mathrm{CMU}}, z_{\mathrm{CMU}}, \rho)
    \end{align*}

    \State Solve:
    $[x_{\mathrm{CMU}}^*, y_{\mathrm{CMU}}^*, z_{\mathrm{CMU}}^*, \rho^*]
    \leftarrow \arg\min_{\mathcal{Y}} f(\mathbf{x})$

    \State Store:
    $P_{\text{CMU},l}^{\text{opt}}(P_t)
    \leftarrow f(x_{\mathrm{CMU}}^*, y_{\mathrm{CMU}}^*, z_{\mathrm{CMU}}^*, \rho^*)$

\EndFor

\State \textbf{Return:} Optimal
$(x^*_{\mathrm{CMU}}, y^*_{\mathrm{CMU}}, z^*_{\mathrm{CMU}})$,
$\rho^*$, and $P_{\text{CMU},l}^{\text{opt}*}(\text{out})$

\end{algorithmic}
\end{algorithm}

{\color{black}
P4 is inherently non-convex because the CMU outage probability depends jointly on the CMU 3D position and the PS factor $\rho$, resulting in a highly nonlinear objective function. In particular, the received signal power used for EH includes a distance-dependent path loss and Nakagami-$m$ fading, determined by the channel between the CMU and the ground IoT devices. Since the channel gain depends nonlinearly on the CMU coordinates, the resulting SINR expressions become nonlinear with respect to the optimization variables. Moreover, the PS factor $\rho$ appears multiplicatively in both the SINR and EH expressions, creating strong coupling between $\rho$ and the received signal power used for both information decoding and EH. Consequently, the objective function is non-separable and exhibits non-convex behavior, making it intractable for conventional convex optimization techniques.}
Unlike \cite{9519666}, which only optimized the horizontal coordinate with a fixed {\color{black}$y_{CMU}$, $z_{CMU}$}, this study aims to jointly optimize the full 3D coordinates along with the PS factor.
To this end, two algorithms are proposed. First, we propose the CPPF method that decomposes the original high-dimensional non-convex problem into simpler sub-problems, which are then solved sequentially. Each variable is optimized separately while holding the other variables constant, and the iteration process is repeated until convergence is reached, as described in Algorithm 2.
In contrast to CPPF, which optimizes one variable at a time while fixing the others (which may lead to a sub-optimal solution), we propose a metaheuristic algorithm, called JCPPM (given in Algorithm 3), that simultaneously optimizes all variables ${\color{black}x_{CMU}, y_{CMU}, z_{CMU}}$, and $\rho$. Although JCPPM is also a sub-optimal method, it provides improved performance over CPPF by jointly optimizing all variables instead of sequentially updating them.

\begin{table}[t]
\caption{{\color{black}Comparison of CPPF and JCPPM Algorithms}}
\color{black}
\centering
\resizebox{\linewidth}{!}{%
\begin{tabular}{|l|l|l|}
\hline
\textbf{Feature} & \textbf{CPPF} & \textbf{JCPPM} \\ \hline
\textbf{Optimization approach}    & Sub-problem decomposition & \textbf{Joint optimization} \\ \hline
\textbf{Computational complexity} & High & Moderate \\ \hline
\textbf{Convergence }             & Slow, sub-optimal & Robust, sub-optimal \\ \hline
\textbf{Robustness }              & Sensitive to initialization & Less sensitive, more robust \\ \hline
\textbf{Performance  }            & Moderate & Higher \\ \hline
\end{tabular}%
}
\label{tab:comparison}
\end{table}

\textbf{{\color{black}Computational Complexity Analysis:}}
The proposed Algorithm 2 has a computational complexity of approximately ${\color{black}\mathcal{O}(I_{x_{CMU}} I_{x_{CMU}} I_{x_{CMU}} I_{\rho})}$, where {\color{black}$I_{x_{CMU}}$, $I_{y_{CMU}}$, $I_{z_{CMU}}$, and $I_\rho$} represent the lengths of {\color{black}$x_{CMU}$, $y_{CMU}$, $z_{CMU}$, and $\rho$}, respectively, and $\mathcal{O}$ denotes the Big-$\mathcal{O}$ notation. Moreover, the computational complexity of Algorithm 3 primarily depends on the number of fitness evaluations, approximately $\mathcal{O}(\mathcal{G} \ \mathcal{P} \ \mathcal{C}_{\mathrm{eval}})$, where $\mathcal{G}$ is the number of generations, $\mathcal{P}$ is the population size, and $\mathcal{C}_{\mathrm{eval}}$ is the cost per evaluation. 
This approach optimizes four variables ${\color{black}(x_{\mathrm{CMU}}, y_{\mathrm{CMU}}, z_{\mathrm{CMU}}, \rho)}$ jointly, resulting in lower computational cost compared to sequential algorithms (i.e., CPPF). While JCPPM may produce sub-optimal (rather than exact) solutions, it is substantially less expensive in practice. Specifically, CPPF has complexity ${\color{black}\mathcal{O}(I_{x_{CMU}} I_{x_{CMU}} I_{x_{CMU}} I_{\rho})}$, whereas JCPPM requires $\mathcal{O}(\mathcal{G} \ \mathcal{P} \ \mathcal{C}_{\mathrm{eval}})$. Hence, JCPPM offers reduced complexity and runtime relative to CPPF.
{\color{black}
In practice, CPPF exhibits moderate efficiency but is associated with very high computational complexity and slow convergence. In contrast, JCPPM is efficient and exhibits moderate complexity and robust convergence.
For clarity, Table~\ref{tab:comparison} summarizes the main features, computational complexity, convergence behavior, robustness,~and performance of the CPPF and JCPPM algorithms.}

\section{Numerical Results} 
{\color{black}In this section, we validate the derived analytical expressions and evaluate the performance of the proposed hierarchical ad hoc UAV network with non-linear EH and RSMA under the effect of HWI and ICSI. We consider equal HWI levels, ICSI, and shape parameters for all nodes as in~\cite{beddiaf2023impact,9991954,babaei2018ber,hoang2020outage}, i.e., $K=k_{l}=k_{n}$, $\sigma_{\varepsilon}^2=\sigma_{\varepsilon_l}^2=\sigma_{\varepsilon_n}^2$, and $m=m_{l} =m_{n}$. We assume that the distances between the CHU–CMUs and CMUs–ground IoT devices are fixed as in~\cite{9519666}. The coordinates of the nodes are given in Table I.
All parameters listed in Table~\ref{tab:3} have already been clearly defined and explained in~Table~\ref{tab:1}.}


\begin{table}[!t]
\centering
\caption{Simulation Parameters \cite{10006695,al2014optimal,mozaffari2015drone,khennoufa2024error,9991954,babaei2018ber,hoang2020outage}.}
\begin{tabular}{ll|ll}
\hline
\textbf{Parameter} & \textbf{Value} & \textbf{Parameter} & \textbf{Value} \\ 
\hline
$q_{\text{CHU}}$ & $(40,\, 20,\, 50)$ & $L$ & $5$  \\
$q_{\text{CMU},l}$ & $(15,\, 25,\, 30)$ & $H$ & $20$ \\
$q_{1}$ & $(65,\, 74,\, 0)$ & $M$ & $2$ \\
$q_{2}$ & $(55,\, 70,\, 0)$ & $P_{\text{th}}$ & ${\color{black}5, 10}$ dBm \\
$\xi_{1}$ , $\xi_{2}$ & $9.6$, $0.28$  & $\zeta_{0} = \zeta_{1}$ & $1$ \\
${\color{black}\sigma_{\varepsilon}^2}$ & {\color{black}0.03} & $\zeta_{2}$ & $0.2$ \\
$\rho$ & $0.1$ & $\eta$ & $0.95$ \\
${\color{black}K}$ & ${\color{black}0.1}$ & ${\color{black}m}$ & ${\color{black}1.5, 2, 2.5}$ \\
\hline
\end{tabular}
\label{tab:3}
\end{table}



\begin{figure}
    \centering
    \includegraphics[width=0.9\linewidth]{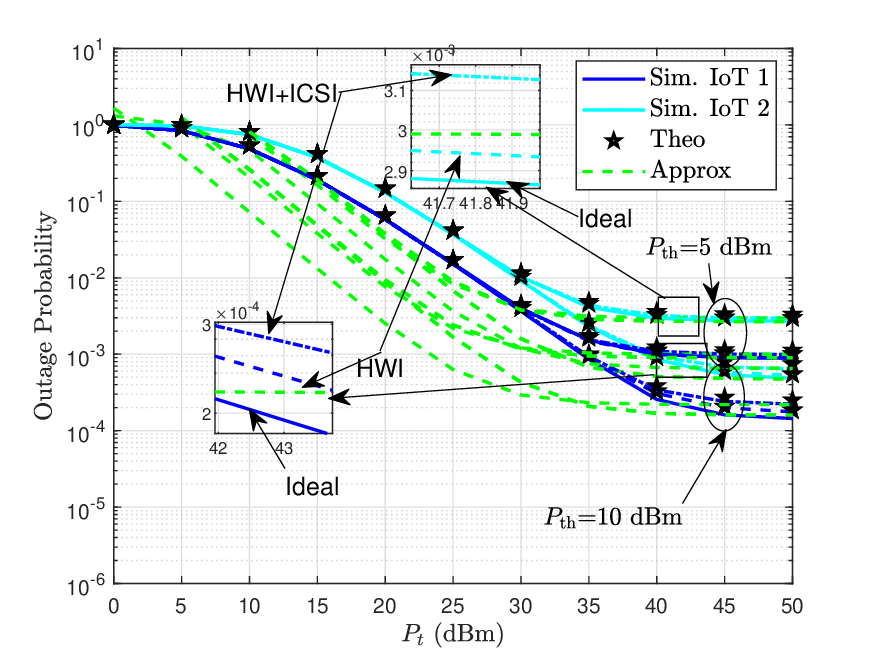}
    \caption{{\color{black} Outage probability performance of two ground IoT devices with different threshold values under HWI and ICSI}.}
    \label{fig:fig3}
\end{figure}

{\color{black}Fig.~\ref{fig:fig3} presents the outage probability performance of two ground IoT devices that are derived from the expression in (15) with different $P_{\text{th}}=\{5, 10\}$ dBm as in~\cite{babaei2022performance}, when HWI factor $K=\{0, 0.1\}$ and ICSI factor $\sigma_{\varepsilon}^2=\{0, 0.03\}$ as in~\cite{9991954}.} The numerical results closely match the simulation results, validating our analytical expressions.
Moreover, the approximate outage probability for ground IoT devices is also obtained. At high transmit power values, the asymptotic outage probability curve of the system is relatively limited over the theoretical curves, demonstrating the expected performance trends at high transmit power levels. 
At low transmit power values, the outage probability decreases rapidly as $P_t$ increases, since higher transmit power enhances the received SINR and improves both information decoding and {\color{black}EH}. However, due to the {\color{black}non-linear} characteristics of practical EH circuits described in (6)–(7), this improvement ceases once the received power $P_t|g_l|^2$ exceeds the saturation threshold $P_{\text{th}}$. Beyond this point, the harvested energy remains constant, and further increases in $P_t$ no longer improve the link quality, resulting in an outage probability floor at high transmit power levels. This behavior reflects the saturation effect of realistic rectifier circuits used in EH modules.
{\color{black}Additionally, non-ideal cases (i.e., HWI and ICSI) have a detrimental effect on outage performance and has a negative impact on the performance of both ground IoT devices. 
It is observed that the impact of HWI is slight, but in the presence of ICSI, the degradation in performance becomes more pronounced. Specifically, the mismatch between the estimated and actual channel coefficients, combined with residual distortion noise introduced by HWI at the transceiver, reduces the effective SINR by both adding interference and diminishing the useful received signal power. This affects the reliability of the decoding and efficiency in harvested energy, leading to an increased outage probability. This again emphasizes the significance of mitigating HWI and ICSI to fully exploit the benefits of EH in the proposed system.}

\begin{figure}
    \centering
    \includegraphics[width=0.9\linewidth]{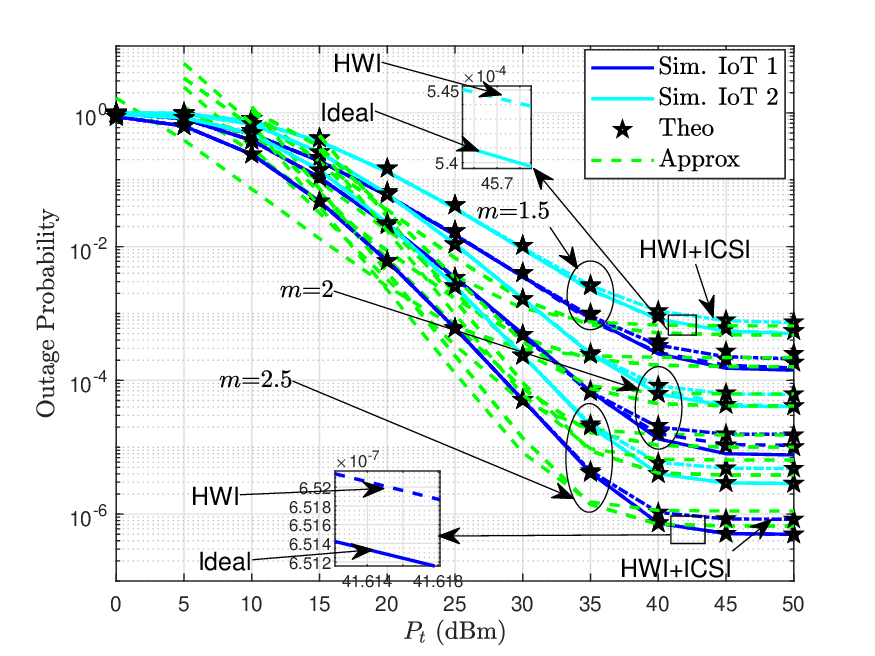}
    \caption{{\color{black}Outage probability performance of two ground IoT devices with different $m$ values, under HWI and ICSI.}}
    \label{fig:fig4}
\end{figure}

Fig.~\ref{fig:fig4} presents the outage probability performance of two ground IoT devices with different shape parameters, $m = \{1.5, 2, 2.5\}$, when $K = 0.3$. In this figure, the analysis refers to both integer and non-integer values of $m$ and demonstrates that the shape parameter $m$ governs the diversity order. We observe that as $m$ increases, the outage probability curves of the ground IoT devices become steeper, indicating enhanced diversity order. A higher $m$ indicates less severe fading in the Nakagami-$m$ model, meaning fewer deep fades and higher average received power, which reduces the outage probability.
{\color{black}Moreover, this figure illustrates the negative impact of HWI and ICSI on outage performance at different values of m. As shown in this figure, the impact of HWI is small, while the addition of ICSI has a considerable negative impact on system performance. Hence, although the fading conditions improve with an increase in m, the presence of HWI and ICSI restricts the diversity gains. Therefore, higher values of m still result in steeper outage curves and improved performance, but the overall diversity order remains constrained by the combined effect of HWI and ICSI.}

\begin{figure}
    \centering
    \includegraphics[width=0.9\linewidth]{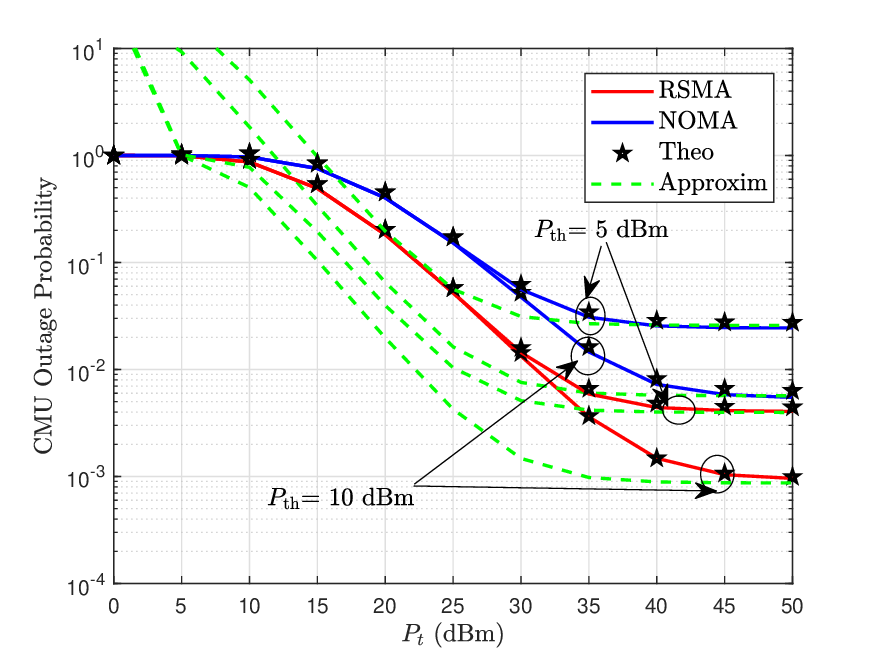}
    \caption{{\color{black}Comparison of the hierarchical ad hoc UAV network with non-linear EH using NOMA and RSMA with different threshold values under HWI and ICSI.}}
    \label{fig:fig5}
\end{figure}

\begin{figure}
    \centering
    \includegraphics[width=0.9\linewidth]{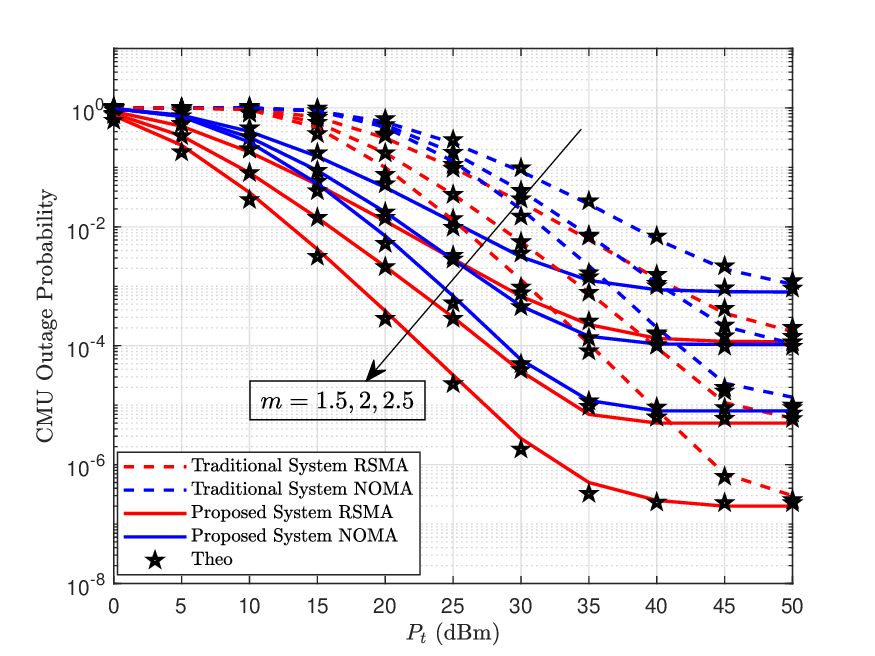}
    \caption{{\color{black}Comparison between the traditional system of ground-based EH and the proposed system under HWI and ICSI.}}
    \label{fig:fig6}
\end{figure}

Moreover, to demonstrate the superiority of our proposed system, we compare it with two schemes: NOMA and ground-based EH (i.e., collecting energy from a ground source) as presented in Fig.~\ref{fig:fig5} and Fig.~\ref{fig:fig6}, respectively. In Fig.~\ref{fig:fig5}, the outage probability performance of the proposed RSMA-based system is compared with NOMA under different EH threshold values {\color{black}($P_{\text{th}}=\{5, 10\}$ dBm) under HWI ($K=0.1$) and ICSI ($\sigma_{\varepsilon}^2=0.03$)}. The figure shows that RSMA outperforms NOMA in hierarchical ad hoc UAV network environments, especially at high transmit powers. This performance gain, approximately 5–8 dB, arises because RSMA splits each user’s message into common and private parts, enabling more flexible and efficient interference management compared to NOMA. Moreover, the approximate results are relatively consistent with theory, demonstrating the reliability of the analytical results. In addition, RSMA maintains its superiority across different $P_{\text{th}}$ values, confirming its robustness under varying EH conditions.
In Fig.~\ref{fig:fig6}, we compare the proposed hierarchical ad hoc UAV system with non-linear EH against a conventional ground-based EH system for both RSMA and NOMA. For a fair comparison, we assume that the distance between the harvested node and the transmitter in both systems (i.e., the proposed system and the ground-based EH system) is 30 m. We observe that the proposed system (i.e., hierarchical ad hoc UAV with non-linear EH) with NOMA and RSMA outperforms the traditional system (i.e., ground-based EH system) by approximately 13 dB and 16 dB, respectively. This improvement results from the favorable A2A propagation characteristics between UAVs, which provide strong LoS links and eliminate NLoS impairments such as buildings or terrain blockage, enabling efficient energy transfer. 
{\color{black}In addition, the figure shows that as the value of $m$ increases ($m = \{1.5, 2, 2.5\}$), the outage performance improves, demonstrating that the proposed system remains robust under varying fading conditions compared to NOMA.}
It can also be seen that RSMA consistently outperforms NOMA in all cases, highlighting its efficiency in managing interference and enhancing both energy and spectral efficiency in UAV-assisted networks.

\begin{figure}[t]
\centering
\subfloat[]{%
    \includegraphics[width=0.9\linewidth]{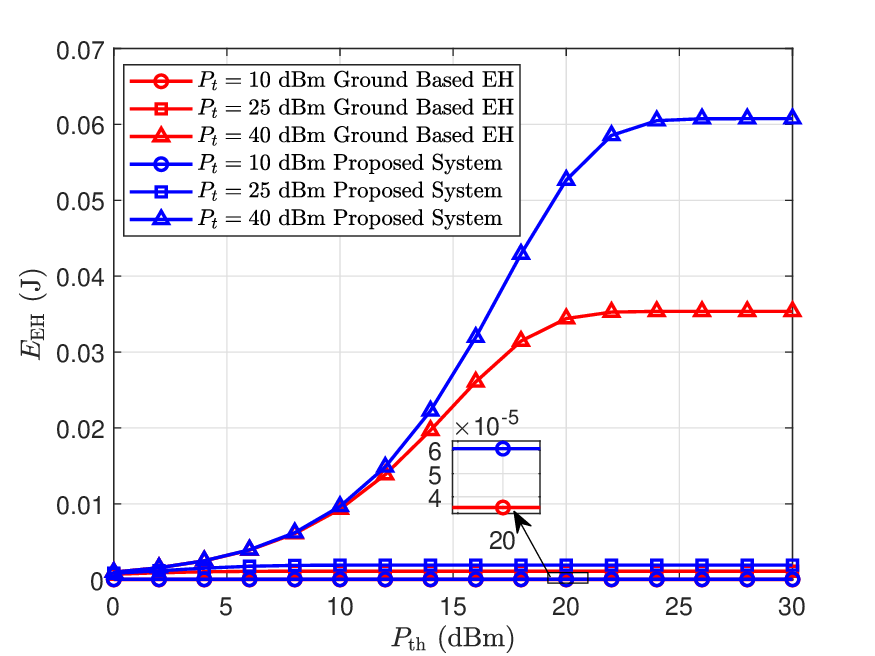}
        \label{fig:fig99a}
}

\vspace{0.5cm}

\subfloat[]{%
    \includegraphics[width=0.9\linewidth]{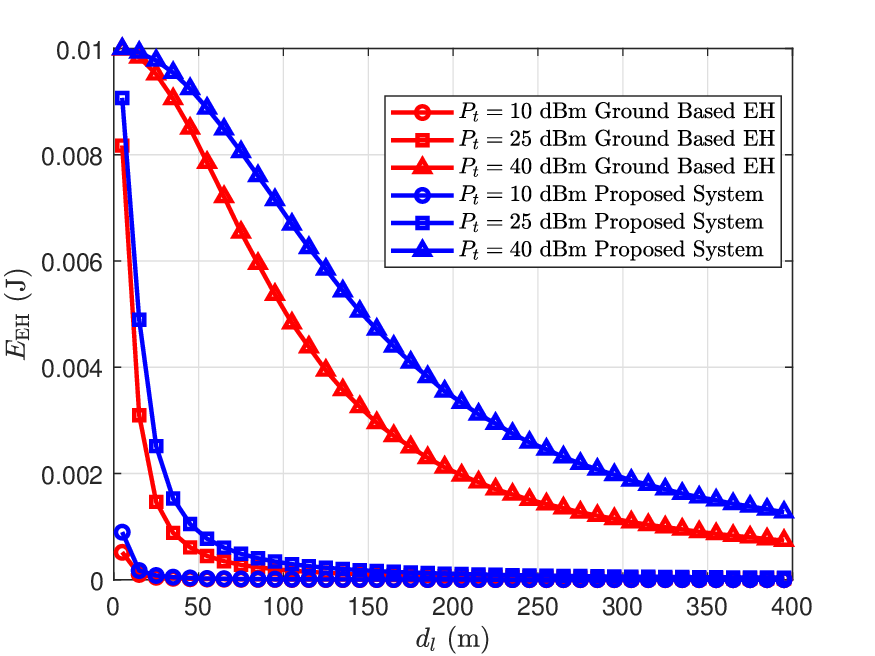}
        \label{fig:fig99b}
}

\caption{{\color{myblue}Comparison of the harvested energy between the ground-based EH and the proposed system; (a) $E_{\text{EH}}$ w.r.t., $P_{\text{th}}$ with different $P_{t}$ values; (b) $E_{\text{EH}}$ w.r.t., $d_{l}$ with different $P_{t}$ values.}}
    \label{fig:fig99}
\end{figure}

To demonstrate the effectiveness of the proposed system in improving EH efficiency compared to a conventional ground-based EH system, Fig.~\ref{fig:fig99} shows the harvested energy of both systems across various parameters. Specifically, Fig.~\ref{fig:fig99a} illustrates the relationship between harvested energy ($E_{\text{EH}}$), the transmit power $P_{\text{t}}$, and the saturation threshold $P_{\text{th}}$. 
{\color{myblue}It is observed that the amount of EH ($E_{\text{EH}}$) increases with both increasing $P_{\text{t}}$ and $P_{\text{th}}$, as higher $P_{\text{t}}$ enhances the received RF energy, while a larger $P_{\text{th}}$ allows the EH circuit to operate in its linear region for a wider range of input power levels. The proposed system consistently achieves higher harvested energy compared to the ground-based EH, particularly at medium and high $P_{\text{t}}$ values (e.g., $P_{\text{t}}=\{25, 40\}$ dBm). This improvement is attributed to the favorable A2A propagation conditions in UAV-assisted environments, which enable stronger LoS links and more effective energy transfer. This confirms the advantage of the proposed system in improving EH performance in UAV environments.
Fig.~\ref{fig:fig99b} further shows the variation of harvested energy with respect to the distance between the CHU and CMU ($d_{l}$) with different $P_{\text{t}}$ values (i.e., $P_{\text{t}}=\{10, 25, 40\}$ dBm). It is observed that EH decreases with increasing distance $d_{l}$, which aligns with the expected path loss behavior. However, the proposed system maintains significantly higher EH levels compared to the ground-based EH system over the considered distance range.  In particular, the proposed system demonstrates improved robustness to distance variations compared to the ground-based EH system, maintaining higher harvested energy levels even as $d_{l}$ increases, especially at medium and high $P_{\text{t}}$ levels (e.g., $P_{\text{t}}=\{25, 40\}$ dBm). This behavior highlights the advantage of the proposed UAV-assisted architecture in maintaining more efficient wireless energy transfer under extended coverage conditions.}
{\color{black}As illustrated in this figure, although this amount of EH during a one-second transmission period is relatively small, it is sufficient to support low-power communication and relay circuits. Compared with ground-based EH, the proposed system achieves relatively higher harvested energy, highlighting its potential for UAV relaying applications. These results motivate further research to integrate advanced techniques to enhance EH and improve the reliability of UAV transmissions.}

\begin{figure}
    \centering
    \includegraphics[width=0.9\linewidth]{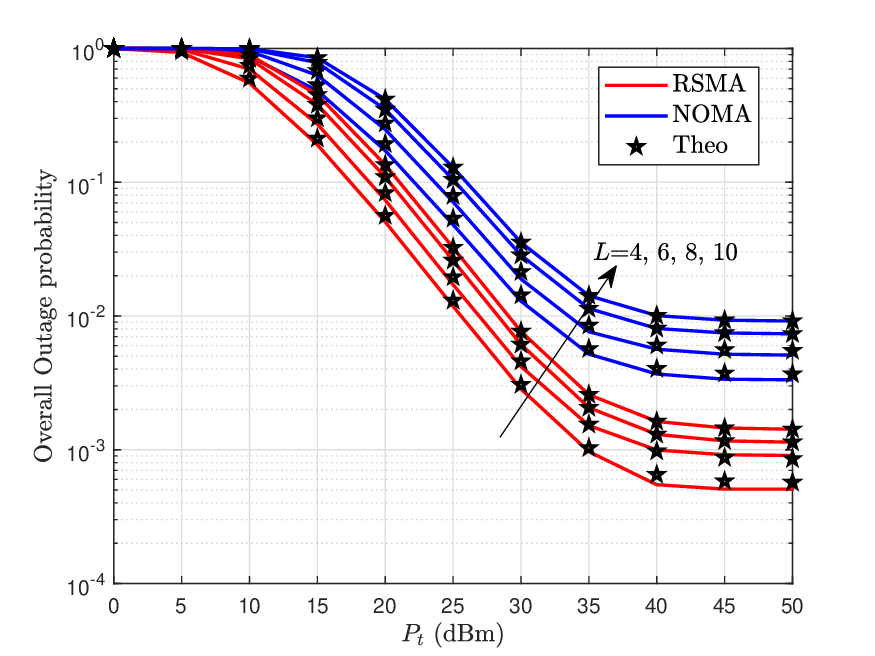}
    \caption{{\color{black}Overall outage probability for RSMA and NOMA with different numbers of CMU, $L=\{4, 6, 8, 10\}$ under HWI and ICSI}.}
    \label{fig:fig7}
\end{figure}

\begin{figure}
    \centering
    \includegraphics[width=0.9\linewidth]{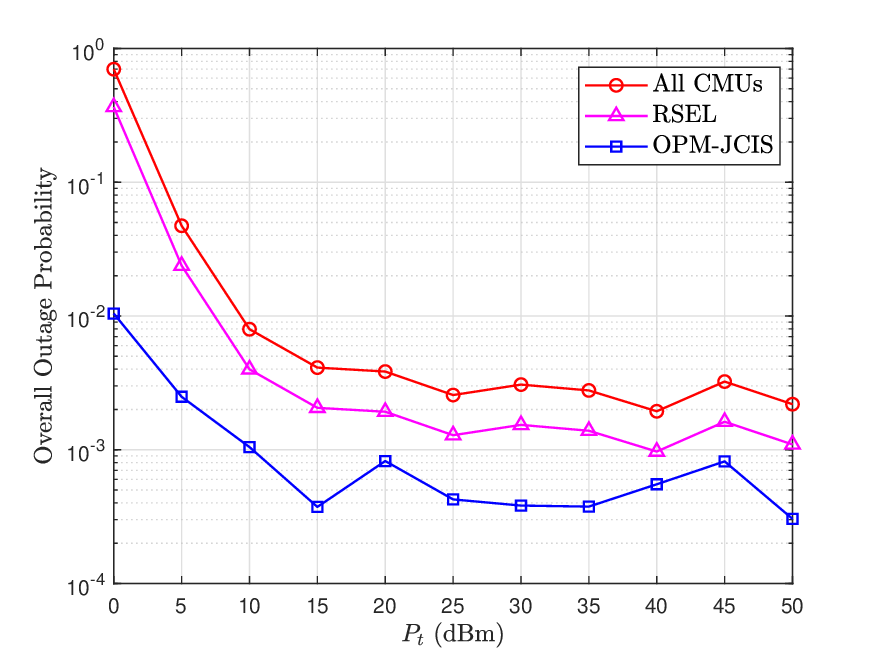}
    \caption{{\color{black}Overall outage probability w.r.t. $P_t$ for various CMU selection algorithms: All CMUs, RSEL, and OPM-JCIS under HWI and ICSI.}}
    \label{fig:fig8}
\end{figure}

Furthermore, to illustrate the trade‑off between system performance and the number of CMUs used to enhance coverage, Fig.~\ref{fig:fig7} shows the overall outage probability for different numbers of CMUs  $L=\{4, 6, 8, 10\}$ using RSMA and NOMA. It can be observed that increasing the number of CMUs degrades the overall outage performance in both RSMA and NOMA, because as more CMUs participate in the relaying process, the system becomes more prone to link failures — the probability that at least one CMU experiences poor channel conditions or outage increases. This happens because adding more CMUs increases the chance that one of them experiences a poor link, which raises the overall outage probability. Therefore, it is essential to carefully consider the trade‑off between system performance and the number of CMUs.
To this end, and to achieve optimal performance and quality of service, we focused on demonstrating the effectiveness of selecting CMUs with optimal conditions and their impact on system performance using the OPM-JCIS algorithm. 
Therefore, to better evaluate the results of CMU selection, Fig.~\ref{fig:fig8} compares the overall outage probability performance (given in (29)) between the proposed OPM-JCIS algorithm, the {\color{black}RSEL} algorithm, and the case with all CMUs active. 
The {\color{black}RSEL} algorithm selects CMUs randomly without considering their channel conditions, while all CMUs' is obtained based on (29) without any algorithm. It can be seen that the proposed OPM-JCIS algorithm achieves higher performance gain compared to the benchmarks (i.e., {\color{black}RSEL} algorithm and all CMUs) across all transmit power values. This yields that the proposed OPM-JCIS algorithm enhances the efficiency of the system by selecting the CMU with the best channel conditions, which increases the received signal quality and reduces the impact of weak relay links, balancing performance gains with system complexity.

\begin{figure}{}
    \centering
    \includegraphics[width=.7\linewidth]{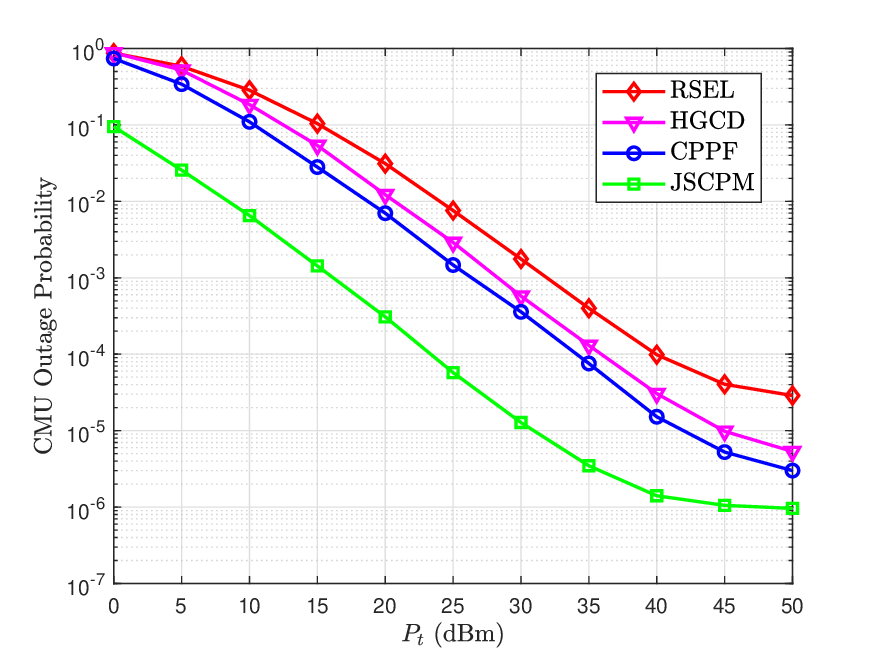}
    \caption{{\color{black}CMU outage probability w.r.t. $P_t$ for various optimization algorithms: RSEL, HGCD, CPPF, and JCPPM under HWI and ICSI.}}
    \label{fig:fig11}
\end{figure}

Furthermore, the CMU position and PS factor ($\rho$) are critical for improving system efficiency. Fig.~\ref{fig:fig11} compares the CMU outage probability performance of different algorithms, such as RSEL, HGCD, CPPF, and JCPPM, under HWI and ICSI. We observe that the CPPF and HGCD algorithms achieve intermediate CMU outage probability performance compared to the {\color{black}RSEL} algorithm, which does not consider the joint selection of the optimal CMU coordinates and the EH factor ($\rho$) due to its random nature. The {\color{black}RSEL} algorithm, on the other hand, performs worst since its random selection cannot guarantee favorable channel or PS conditions. In contrast, the JCPPM algorithm achieves the lowest CMU outage probability among all algorithms, because it jointly optimizes both the CMU coordinates and the PS factor ($\rho$), enabling simultaneous enhancement of signal reception and EH. {\color{black}This optimization process provides a balanced trade-off between path loss reduction and harvested energy maximization, thereby ensuring the availability of adequate amounts of energy for sustainable operation and signal decoding, thereby improving outage performance significantly.}

\section{Conclusion}
{\color{black}
This study proposed a hierarchical ad hoc UAV network with non-linear EH and RSMA to enhance both energy and cost efficiency. For a practical scenario, the effects of HWI and ICSI were accounted for in the proposed system. We examined the outage probability regarding ground IoT devices, each CMU, and the overall outage probability of the proposed system over Nakagami-$m$ fading channels, accounting for practical challenges, such as non-linear EH, HWI, and ICSI.  Additionally, approximate outage probability expressions are derived for high transmit power regimes. Moreover, two optimization problems were formulated to enhance communication reliability: one aims to minimize outage probability by optimizing ground IoT device pairing and CMU selection, while the second determines optimal CMU hovering coordinates and EH factors. 
The proposed system and algorithms were evaluated against multiple benchmarks (NOMA, ground-based EH, all CMUs, RSEL, and HGCD). The findings revealed that RSMA outperforms NOMA in hierarchical ad hoc UAV networks with non-linear EH, achieving approximately 3–4 dB gain. Additionally, the proposed UAV-based EH framework outperformed traditional ground-based EH, with gains of approximately 8 dB and 12 dB in NOMA and RSMA, respectively. The proposed algorithms (OPM-JCIS, CPPF, and JCPPM) further deliver superior performance over all benchmarks. Overall, optimized CMU selection, IoT pairing, and EH factor tuning significantly improve system efficiency while maintaining a balance between performance and complexity. The impact of HWI, ICSI, and the non-linearity of EH circuits reduced the overall performance of the system, highlighting the need for future research on effective mitigation techniques.
The findings underscore that reducing outage probability does not equate to maximizing EH; instead, the optimal design balances EH efficiency and communication reliability. Future work will extend this model to consider mobility-induced Doppler effect, HWI mitigation techniques, and machine learning-based optimization for real-time trajectory and resource management.}

\appendices 




\section{Proof of Theorem 1}
At the second $\sfrac{T}{2}$ time slot, the ground IoT devices decode the common message $x_{,\mathrm{c}}$ by regarding the private messages of the ground IoT devices as interference or noise. After successful decoding of $x_{,\mathrm{c}}$, SIC is used to remove the decoded $x_{,\mathrm{c}}$ from the received signal. The private signal of IoT devices, $x_{,\mathrm{p},n}$, is decoded by treating the private signals of other ground IoT devices as interference. Hence, the outage probability of the second $\sfrac{T}{2}$ time slot at the ground IoT devices can be expressed as
\begin{equation}
    P_{n}^{II} (\text{out})= 1-\Pr \left(  \gamma_{n,\mathrm{c}} > \varphi_{n,\mathrm{c}}, \gamma_{n,\mathrm{p}} > \varphi_{n} \right).
\end{equation}

Thus, (36) can be rewritten as \cite{beddiaf2023impact}
\begin{equation}
\begin{split}
&    P_{n}^{II}  (\text{out})=\\& 1- \prod_{n}^{M}  \left(1- \underbrace{ \Pr \left(  \gamma_{n,\mathrm{c}} \leq \varphi_{n,\mathrm{c}}\right) }_{\mathcal{A}} \right)  \left(1-\underbrace{\Pr \left( \gamma_{n,\mathrm{p}} \leq \varphi_{n} \right)}_{\mathcal{B}} \right).
\end{split}
\end{equation}


Through (37), $\mathcal{A}$ and $\mathcal{B}$ define $\gamma_{n,\mathrm{c}}$ and $\gamma_{n,\mathrm{p}}$ under two conditions depending on $P_{t} |{g}_{l}|^2$ relative to $P_{\text{th}}$. This implies the outage probability computation in (37) will be piecewise, based on whether it is above or below $P_{\text{th}}$. Thus, after some algebraic simplifications, (37) can be expressed as
\begin{equation}\label{eq:17}
\begin{split}
& P_{\text{out},n}^{II}=  1- \prod_{n}^{M}  \ \left(1- \left( \mathcal{I}_1 + \mathcal{I}_2 \right) \right) \times \left(1-\left( \mathcal{I}_3 + \mathcal{I}_4 \right) \right),
\end{split}
\end{equation}
where \begin{math}
   \mathcal{I}_1=  \Pr  ( |{g}_{n}|^2 |{g}_{l}|^2 \leq \mho_{n,\mathrm{c}}^{A}, P_{t} |{g}_{l}|^2 \leq P_{\text{th}} )
\end{math}, 
\begin{math}
   \mathcal{I}_2= \Pr  ( |{g}_{n}|^2 \leq \mho_{n,\mathrm{c}}^{B}, P_{t} |{g}_{l}|^2 > P_{\text{th}} ) 
\end{math}, 
\begin{math} \mathcal{I}_3=  \Pr  ( |{g}_{n}|^2 |{g}_{l}|^2 \leq \mho_{n,\mathrm{p}}^{A}, P_{t} |{g}_{l}|^2 \leq P_{\text{th}} ) \end{math}, and 
\begin{math}
   \mathcal{I}_4=  \Pr ( |{g}_{n}|^2  \leq \mho_{n,\mathrm{p}}^{A}, P_{t} |{g}_{l}|^2 > P_{\text{th}} )
\end{math}. We compute each term of (38), respectively, below. $\mathcal{I}_1$ can be written as
\begin{equation}
   \mathcal{I}_1=  \Pr  ( |{g}_{n}|^2  \leq \frac{\mho_{n,\mathrm{c}}^{A}}{|{g}_{l}|^2},  |{g}_{l}|^2 \leq \frac{P_{\text{th}}}{P_{t}} ).
\end{equation}

To calculate (39), we apply the law of total probability, integrating over the distribution of $|{g}_{l}|^2$ and conditioning on its value. This produces the following integral expression
\begin{equation}
   \mathcal{I}_1= \int_{0}^{\frac{P_{th}}{P_{t}}} F_{|{g}_{l}|^2} \left( \frac{\mho_{n,c}^{A}}{x} \right) f_{|{g}_{n}|^2}(x) \mathrm{d}x.
\end{equation}

Substituting the CDF of $|{g}_{l}|^2$ and the PDF of $|{g}_{n}|^2$, assuming independent Nakagami-$m$ fading channels, yields the expression
\begin{equation}
    \mathcal{I}_1 = \int_{0}^{\frac{P_{th}}{P_{t}}} \frac{\Upsilon\left( {\color{black}m_{l}}, \frac{ {\color{black}m_{l}} \mho_{n,c}^{A}}{\Omega_{l} x}\right)}{\Gamma( {\color{black}m_{l}})} 
    \frac{m_{n}^{m_{n}} x^{m_{n} - 1}}{\Gamma(m_{n}) \Omega_{n}^{m_{n}}} e^{-\frac{m_{n} }{\Omega_{n}}x} \mathrm{d}x.
\end{equation}

This integral cannot be obtained in closed form based on the authors’ knowledge. However, it can be easily determined by using numerical tools.

In similar way, the second component $\mathcal{I}_2$ of (38) can be given by
\begin{equation}
\begin{split}
 &  \mathcal{I}_2= \Pr   \left( |{g}_{n}|^2 \leq \mho_{n,c}^{B}\right) 
 \Pr \left( |{g}_{l}|^2 > \frac{P_{\text{th}}}{P_{t}} \right)  .
\end{split}
\end{equation}

By employing (13) and (14) of the dedicated Nakagami-$m$ fading channel and after some algebraic manipulation, (42) is obtained as
\begin{equation}
\begin{split}
 &  \mathcal{I}_2= 
 \frac{\Upsilon(m_{n},\frac{m_{n}}{\Omega_{n}}\mho_{n,c}^{B})}{\Gamma(m_{n})} - \frac{\Upsilon(m_{n},\frac{m_{n}}{\Omega_{n}}\mho_{n,c}^{B})}{\Gamma(m_{n})} 
 \frac{\Upsilon(m_{l},\frac{m_{l}}{\Omega_{l}}\chi)}{\Gamma(m_{l})}   .
\end{split}
\end{equation}

Likewise, the third component $\mathcal{I}_3$ of (38) can be expressed by
\begin{equation}
   \mathcal{I}_3=  \Pr  ( |{g}_{n}|^2  \leq \frac{\mho_{n,\mathrm{p}}^{A}}{|{g}_{l}|^2},  |{g}_{l}|^2 \leq \frac{P_{\text{th}}}{P_{t}} )
\end{equation}

Accordingly, (44) can be computed as
\begin{equation}
   \mathcal{I}_3= \int_{0}^{\frac{P_{th}}{P_{t}}} F_{{\color{black}|{g}_{l}|^2}} \left( \frac{\mho_{n,\mathrm{p}}^{A}}{x} \right) f_{|{g}_{n}|^2}(x) \mathrm{d}x.
\end{equation}

Assuming independent Nakagami-$m$ fading channels, and by substituting the CDF of $|{g}_{l}|^2$ and PDF of $|{g}_{n}|^2$, (45) can be obtained 
\begin{equation}
    \mathcal{I}_3 = \int_{0}^{\frac{P_{th}}{P_{t}}} \frac{\Upsilon\left(m_{l}, \frac{m_{l} \mho_{n,\mathrm{p}}^{A} }{\Omega_{l} x}\right)}{\Gamma(m_{l})} 
    \frac{m_{n}^{m_{n}} x^{m_{n} - 1}}{\Gamma(m_{n}) \Omega_{n}^{m_{n}}} e^{-\frac{m_{n} }{\Omega_{n}}x} \mathrm{d}x.
\end{equation}

This integral cannot be obtained in closed form based on the authors’ knowledge. However, it can be efficiently evaluated using numerical tools.

Similarly, the fourth component $\mathcal{I}_4$ of (38) can be provided~by
\begin{equation}
\begin{split}
 &  \mathcal{I}_4= \Pr   \left( |{g}_{n}|^2 \leq \mho_{n,\mathrm{p}}^{B}\right)
 \left( |{g}_{l}|^2 > \frac{P_{\text{th}}}{P_{t}} \right)  .
\end{split}
\end{equation}

By using (13) and (14), along with some algebraic manipulation, (47) can be expressed as
\begin{equation}
\begin{split}
 &  \mathcal{I}_4= 
 \frac{\Upsilon(m_{n},\frac{m_{n}}{\Omega_{n}}\mho_{n,\mathrm{p}}^{B})}{\Gamma(m_{n})} - \frac{\Upsilon(m_{n},\frac{m_{n}}{\Omega_{n}}\mho_{n,\mathrm{p}}^{B})}{\Gamma(m_{n})} 
 \frac{\Upsilon(m_{l},\frac{m_{l}}{\Omega_{l}}\chi)}{\Gamma(m_{l})}   .
\end{split}
\end{equation}

By substituting (41), (43), (46), and (48) into (38), we obtain the outage probability as in (17).
The proof is completed.

\section{Proof of Corollary 1}

{\color{black}
To gain further insight into the system performance, the approximate outage probability is presented for the case when $P_{t} \rightarrow  \infty$. To get an approximation of the outage probability of the second $\sfrac{T}{2}$ time slot, we calculate the approximation of each term of (38) as detailed below. The approximate outage probability of $\mathcal{I}_1$ can be expressed as
\begin{equation}
    \mathcal{I}_1^{\infty}  \approx \int_{0}^{\frac{P_{\text{th}}}{P_{t}}} \frac{\left(\frac{m_{l} \mho_{n,c}^{A} }{\Omega_{l} }\right)^{m_{l}} }{m_{l} \Gamma(m_{l})} \cdot \frac{m_{n}^{m_{n}} x^{m_{n}-m_{l} - 1}}{\Gamma(m_{n}) \Omega_{n}^{m_{n}}} e^{-\frac{m_{n} }{\Omega_{n}}x} \mathrm{d}x .
\end{equation}

Hence, after some algebraic manipulation, (49) can be rewritten as follows
\begin{equation}
    \mathcal{I}_1^{\infty}  \approx \int_{0}^{\frac{P_{\text{th}}}{P_{t}}} \frac{\left(  1 
    +  \frac{\mho_{n,\mathrm{c}}^{A_2}}{\mho_{n,\mathrm{c}}^{A_1}} x  \right)^{m_{l}} ( \mho_{n,\mathrm{c}}^{A_1})^{m_{l}} }{m_{l} \Gamma(m_{l})} \cdot \frac{m_{n}^{m_{n}} x^{m_{n}-m_{l} - 1}}{\Gamma(m_{n}) \Omega_{n}^{m_{n}}} e^{-\frac{m_{n} }{\Omega_{n}}x} \mathrm{d}x .
\end{equation}

For high transmit power (i.e., $\frac{\mho_{n,\mathrm{c}}^{A_2}}{\mho_{n,\mathrm{c}}^{A_1}} x \rightarrow 0$), using the first-order binomial expansion for approximation as defined in~\cite{morales1989generalization}, (50) can be rewritten after some algebraic manipulation as 
\begin{equation}
\begin{split}
   & 
    \mathcal{I}_1^{\infty}  \approx \frac{ m_{n}^{m_{n}} \left(   \mho_{n,\mathrm{c}}^{A_1}   \right)^{m_{l}}  }{m_{l} \Gamma(m_{l}) \Gamma(m_{n}) \Omega_{n}^{m_{n}}} 
    \underbrace{\int_{0}^{\frac{P_{\text{th}}}{P_{t}}}  
     x^{m_{n}-m_{l} - 1} e^{-\frac{m_{n} }{\Omega_{n}}x} \mathrm{d}x }_{\mathcal{C}}
   \\& + \frac{ m_{n}^{m_{n}}   \mho_{n,\mathrm{c}}^{A_2} (\mho_{n,\mathrm{c}}^{A_1})^{m_{l}-1}     }{ \Gamma(m_{n}) \Omega_{n}^{m_{n}} \Gamma(m_{l})}
    \underbrace{ \int_{0}^{\frac{P_{\text{th}}}{P_{t}}} 
     x^{m_{n}-m_{l} } e^{-\frac{m_{n} }{\Omega_{n}}x} \mathrm{d}x}_{\mathcal{B}}.
    \end{split}
\end{equation}



To solve this integral (i.e., $\mathcal{C}$ and $\mathcal{B}$), let us execute a substitution to align with the form of the incomplete gamma function. Let:
\begin{math}
    t = \frac{m_n}{\Omega_n} x \quad \Rightarrow \quad \mathrm{d}t = \frac{m_n}{\Omega_n} \mathrm{d}x \quad \Rightarrow \quad \mathrm{d}x = \frac{\Omega_n}{m_n} \mathrm{d}t.
\end{math}
When $x = 0$, $t = 0$, and when $x = \frac{P_{\text{th}}}{P_{t}}$, $t = \frac{m_n P_{\text{th}}}{\Omega_n P_{t}}$. By substituting into the integral of $\mathcal{C}$ and after some algebraic manipulation, we obtain
\begin{equation}
   \mathcal{C} =\left(\frac{\Omega_n}{m_n} \right)^{m_n - m_l}  \int_{0}^{\frac{m_n P_{\text{th}}}{\Omega_n P_{t}} }  {t}^{m_n - m_l - 1}  e^{-t}   \mathrm{d}t.
\end{equation}

This integral can be computed in the form of the lower incomplete gamma function as
\begin{equation}
   \mathcal{C} = \left(\frac{\Omega_n}{m_n}\right)^{m_n - m_l} \Upsilon\left(m_n - m_l, \frac{m_n P_{\text{th}}}{\Omega_n P_{\text{t}}} \right).
\end{equation}

Likewise, after some algebraic manipulation, we obtain the integral of $\mathcal{B}$ as
\begin{equation}
   \mathcal{B} =\left(\frac{\Omega_n}{m_n} \right)^{m_n - m_l+1}  \int_{0}^{\frac{m_n P_{\text{th}}}{\Omega_n P_{t}} }  {t}^{m_n - m_l }  e^{-t}   \mathrm{d}t.
\end{equation}

This integral can be computed in the form of the lower incomplete gamma function as
\begin{equation}
   \mathcal{B} = \left(\frac{\Omega_n}{m_n}\right)^{m_n - m_l+1} \Upsilon\left(m_n - m_l +1, \frac{m_n P_{\text{th}}}{\Omega_n P_{\text{t}}} \right).
\end{equation}

By substituting (53) and (55) into (51), and after some simplification, we obtain
\begin{equation}
\begin{split}
   & 
    \mathcal{I}_1^{\infty}  \approx 
    \frac{ m_{n}^{m_{n}} \left(   \mho_{n,\mathrm{c}}^{A_1}   \right)^{m_{l}}  }{m_{l} \Gamma(m_{l}) \Gamma(m_{n}) \Omega_{n}^{m_{n}}}  \left(\frac{\Omega_n}{m_n}\right)^{m_n - m_l}
    \\& \left( \mathfrak{V_1} \frac{  \left(   \mho_{n,\mathrm{c}}^{A_1}   \right)^{m_{l}}  }{m_{l} } +\mathfrak{V_2} \frac{ \mho_{n,\mathrm{c}}^{A_2} }{ \mho_{n,\mathrm{c}}^{A_1} } 
    \frac{\Omega_n}{m_n}
      \right),
    \end{split}
\end{equation}
where 
$\mathfrak{V_1}= 
    \Upsilon\left(m_n - m_l, \frac{m_n P_{\text{th}}}{\Omega_n P_{\text{t}}} \right)$ and $
\mathfrak{V_2}= \Upsilon\left(m_n - m_l +1, \frac{m_n P_{\text{th}}}{\Omega_n P_{\text{t}}} \right)$.
}    

Likewise, the second component $\mathcal{I}_2$ of (38) can be approximated as
\begin{equation}
\begin{split}
 &   \mathcal{I}_2^{\infty}  \approx 
 \frac{ (\frac{m_{n}}{\Omega_{n}}\mho_{n,c}^{B})^{m_{n}} }{\Gamma(m_{n}) m_{n}} - \frac{ (\frac{m_{n}}{\Omega_{n}}\mho_{n,c}^{B})^{m_{n}} }{\Gamma(m_{n}) m_{n}} 
 \frac{(\frac{m_{l}}{\Omega_{l}}\chi)^{m_{l}}}{\Gamma(m_{l}) m_{l}}   .
\end{split}
\end{equation}

{\color{black}
By using a similar method to (53), we obtain the approximation of $\mathcal{I}_3$ as
\begin{equation}
\begin{split}
   & 
    \mathcal{I}_3^{\infty}  \approx 
    \frac{ m_{n}^{m_{n}} \left(   \mho_{n,\mathrm{p}}^{A_1}   \right)^{m_{l}}  }{m_{l} \Gamma(m_{l}) \Gamma(m_{n}) \Omega_{n}^{m_{n}}}  \left(\frac{\Omega_n}{m_n}\right)^{m_n - m_l}
    \\& \left( \mathfrak{V_1} \frac{  \left(   \mho_{n,\mathrm{p}}^{A_1}   \right)^{m_{l}}  }{m_{l} } +\mathfrak{V_2} \frac{ \mho_{n,\mathrm{p}}^{A_2} }{ \mho_{n,\mathrm{c}}^{A_1} } 
    \frac{\Omega_n}{m_n}
      \right),
    \end{split}
\end{equation}
}


Similarly, the approximation of the fourth component $\mathcal{I}_4$ of (38) can be provided by
\begin{equation}
\begin{split}
 &  \mathcal{I}_4^{\infty} \approx
 \frac{ (\frac{m_{n}}{\Omega_{n}}\mho_{n,\mathrm{p}}^{B})^{m_{n}} }{\Gamma(m_{n}) m_{n}} - \frac{ (\frac{m_{n}}{\Omega_{n}}\mho_{n,\mathrm{p}}^{B})^{m_{n}} }{\Gamma(m_{n}) m_{n}} 
 \frac{(\frac{m_{l}}{\Omega_{l}}\chi).^{m_{l}}}{\Gamma(m_{l}) m_{l}}   .
\end{split}
\end{equation}

By substituting (56), (57), (58), and (59) into (38), we obtain the approximation of the outage probability at the second $\sfrac{T}{2}$ time slot as given in (23). The proof is completed.


-






\bibliographystyle{IEEEtran}
\bibliography{references}

@ARTICLE{9991954,
  author={Beddiaf, Safia and Khelil, Abdellatif and Khennoufa, Faical and Kara, Ferdi and Kaya, Hakan and Li, Xingwang and Rabie, Khaled and Yanikomeroglu, Halim},
  journal={IEEE Access}, 
  title={{A unified performance analysis of cooperative NOMA with practical constraints: hardware impairment, imperfect SIC and CSI}}, 
  year={2022},
  volume={10},
  number={},
  pages={132931-132948},
  doi={10.1109/ACCESS.2022.3230650}
}

@inproceedings{studer2010mimo,
  title={{MIMO transmission with residual transmit-RF impairments}},
  author={Studer, Christoph and Wenk, Markus and Burg, Andreas},
  booktitle={Internat. ITG Workshop Smart Antennas},
  pages={189--196},
  year={2010},
}

@ARTICLE{10836898,
  author={Ning, Bing and Yi, Mengshi and Xu, Yongjun and Li, Jianjun and Hao, Wanming},
  journal={IEEE Trans. Veh. Tech}, 
  title={{UAV-assisted AmBC network: performance analysis in nakagami-$m$ fading channel}}, 
  year={2025},
  volume={74},
  number={5},
  pages={6966-6978},
  doi={10.1109/TVT.2025.3528011}
}

@ARTICLE{10006695,
  author={Yang, Xincheng and Han, Bing and Zhang, Gengxin and Zheng, Ping and Bai, Jianan and Qin, Danyang},
  journal={IEEE Sens. J}, 
  title={{NOMA-assisted routing algorithm design for UAV ad hoc relay networks}}, 
  year={2023},
  volume={23},
  number={3},
  pages={3296-3312},
  doi={10.1109/JSEN.2022.3232636}
}

@ARTICLE{10500741,
  author={Guidotti, Alessandro and Vanelli-Coralli, Alessandro and Jaafari, Mohamed El and Chuberre, Nicolas and Puttonen, Jani and Schena, Vincenzo and Rinelli, Giuseppe and Cioni, Stefano},
  journal={IEEE Access}, 
  title={{Role and evolution of non-terrestrial networks toward 6G systems}}, 
  year={2024},
  volume={12},
  number={},
  pages={55945-55963},
  doi={10.1109/ACCESS.2024.3389459}
}

@ARTICLE{10938203,
  author={Ozturk, Metin and Salamatmoghadasi, Maryam and Yanikomeroglu, Halim},
  journal={IEEE Network}, 
  title={{Integrating terrestrial and non-terrestrial networks for sustainable 6G operations: a latency-aware multi-tier cell-switching approach}}, 
  year={2026},
  volume={40},
  number={2},
  pages={156-164},
  doi={10.1109/MNET.2025.3554393}
}

@ARTICLE{10679214,
  author={Wei, Fengsheng and Feng, Gang and Qin, Shuang and Peng, Youkun and Liu, Yijing},
  journal={IEEE J. Sel. Areas Commun.}, 
  title={{Hierarchical network slicing for UAV-assisted wireless networks with deployment optimization}}, 
  year={2024},
  volume={42},
  number={12},
  pages={3705-3718},
  doi={10.1109/JSAC.2024.3459055}
}

@ARTICLE{10379154,
  author={Pan, Wu and Lv, Na and Hou, Bei and Ren, Zhiyuan},
  journal={IEEE Trans. Net. Sci. Eng.}, 
  title={{Resource allocation and outage probability optimization method for multi-hop UAV relay network for servicing heterogeneous users}}, 
  year={2024},
  volume={11},
  number={3},
  pages={2769-2781},
}

@ARTICLE{9707780,
  author={Nguyen, Anh-Nhat and Ha, Dac-Binh and Vo, Van Nhan and Truong, Van-Truong and Do, Dinh-Thuan and So-In, Chakchai},
  journal={IEEE Access}, 
  title={{Performance analysis and optimization for IoT mobile edge computing networks with RF energy harvesting and UAV relaying}}, 
  year={2022},
  volume={10},
  number={},
  pages={21526-21540},
  doi={10.1109/ACCESS.2022.3150046}
}

@ARTICLE{10930451,
  author={Dong, Chao and Liao, Yiyang and Jia, Ziye and Wu, Qihui and Zhang, Lei},
  journal={IEEE Internet of Things J.}, 
  title={{Joint ADS-B in B5G for hierarchical UAV networks: performance analysis and MEC based optimization}}, 
  year={2025},
  volume={12},
  number={12},
  pages={22211-22223},
  doi={10.1109/JIOT.2025.3552201}
}

@ARTICLE{10852192,
  author={Hu, Xiaoyan and Wen, Pengle and Xiao, Han and Wang, Wenjie and Wong, Kai-Kit},
  journal={IEEE Trans. Veh. Tech.}, 
  title={{Maximizing energy charging for UAV-assisted MEC systems with SWIPT}}, 
  year={2025},
  volume={74},
  number={5},
  pages={8442-8447},
  doi={10.1109/TVT.2025.3530426}
}

@ARTICLE{10902111,
  author={He, Yixin and Huang, Fanghui and Wang, Dawei and Zhang, Ruonan},
  journal={IEEE Trans. Net. Sci. Eng.}, 
  title={{Outage probability analysis of MISO-NOMA downlink communications in UAV-assisted Agri-IoT with SWIPT and TAS enhancement}}, 
  year={2025},
  volume={12},
  number={3},
  pages={2151-2164},
  doi={10.1109/TNSE.2025.3545148}
}

@article{CAGIRAN2025155713,
title = {{Outage performance of UAV-assisted relaying MIMO-NOMA networks with SWIPT in Rician fading channels}},
journal = {AEÜ Int. J. Electron. Commun.},
volume = {193},
pages = {155713},
year = {2025},
issn = {1434-8411},
author = {Harun Çağiran and Oğuz Kucur and Saliha Büyükçorak},
}

@ARTICLE{10928333,
  author={Bajpai, Rahul and Shaikh, Afraj Shabbir and Gupta, Naveen},
  journal={IEEE Access}, 
  title={{Wireless powered NOMA assisted full-duplex cooperative U2U communications system with fluctuating two-ray channel}}, 
  year={2025},
  volume={13},
  number={},
  pages={54069-54079},
  doi={10.1109/ACCESS.2025.3551909}
}

@ARTICLE{ahmed2025toward,
  author={Ahmed, Manzoor and Soofi, Aized Amin and Khan, Feroz and Raza, Salman and Khan, Wali Ullah and Su, Lina and Xu, Fang and Han, Zhu},
  journal={IEEE Internet of Things J.}, 
  title={{Toward a sustainable low-altitude economy: A survey of energy-efficient RIS–UAV networks}}, 
  year={2025},
  volume={12},
  number={24},
  pages={51951-51975},
  doi={10.1109/JIOT.2025.3618483}
  }

@ARTICLE{9913422,
  author={Özyurt, S. and Coşkun, A. F. and Büyükçorak, S. and Karabulut Kurt, G. and Kucur, O.},
  journal={IEEE Access}, 
  title={{A survey on multiuser SWIPT communications for 5G+}}, 
  year={2022},
  volume={10},
  number={},
  pages={109814-109849},
  doi={10.1109/ACCESS.2022.3212774}
}

@ARTICLE{9478941,
  author={Sharma, Pankaj Kumar and Gupta, Deepika},
  journal={IEEE Syst. J.}, 
  title={{Outage performance of multi-UAV relaying-based imperfect hardware hybrid satellite-terrestrial networks}}, 
  year={2022},
  volume={16},
  number={2},
  pages={2311-2314},
  doi={10.1109/JSYST.2021.3090799}
}

@ARTICLE{10840246,
  author={Fang, Chao and Feng, Yanxiang and Li, Xiaoling and Yang, Yikang},
  journal={IEEE Trans. Veh. Tech.}, 
  title={{Multi-AAV energy-efficient detection coverage under jamming environment: a hierarchical collaborative learning approach}}, 
  year={2025},
  volume={74},
  number={5},
  pages={7351-7363},
  doi={10.1109/TVT.2025.3529036}
}

@article{lau2023general,
  title={{General outage probability model for UAV-to-UAV links in multi-UAV networks}},
  author={Lau, Wei Jian and Lim, Joanne Mun-Yee and Chong, Chun Yong and Ho, Nee Shen and Ooi, Thomas Wei Min},
  journal={Comput. Net.},
  volume={229},
  pages={109752},
  year={2023},
  publisher={Elsevier}
}

@inproceedings{cui2023distributionally,
  title={{Distributionally robust chance-constrained optimization for hierarchical UAV-based MEC}},
  author={Cui, Can and Jia, Ziye and Dong, Chao and Ling, Zhuang and You, Jiahao and Wu, Qihui},
  booktitle={IEEE Conf. Com. Commun. Workshops},
  pages={1--6},
  year={2023}
}

@article{ma2023vision,
  title={{Vision-based formation control for an outdoor UAV swarm with hierarchical architecture}},
  author={Ma, Liqun and Meng, Dongyuan and Huang, Xu and Zhao, Shuaihe},
  journal={IEEE Access},
  volume={11},
  pages={75134--75151},
  year={2023},
  publisher={IEEE}
}

@ARTICLE{10659004,
  author={Huang, Jie and Yu, Tao and Zhu, Xiaogang and Yang, Fan and Lai, Xianzhi and Alfarraj, Osama and Yu, Keping},
  journal={IEEE Trans. Intell. Transp. Syst.}, 
  title={{Energy efficiency maximization in UAV-assisted intelligent autonomous transport system for 6G networks with energy harvesting}}, 
  year={2025},
  volume={26},
  number={10},
  pages={17212-17222},
  doi={10.1109/TITS.2024.3445088}
}

@ARTICLE{10806653,
  author={Khalaf, Qasim M. and Sali, Aduwati and Ismail, Alyani and Ahmad, Mohd Yazed and Hussein, Yaseein Soubhi and Shah, Jawad Ali},
  journal={IEEE Access}, 
  title={{Advanced nonlinear optimization techniques in SWIPT: a review}}, 
  year={2024},
  volume={12},
  number={},
  pages={192743-192766},
  doi={10.1109/ACCESS.2024.3519522}
}

@article{chen2024performance,
  title={{Performance analysis of UAV-assisted DF relaying network with hardware impairments and energy harvesting}},
  author={Chen, Jielin and Chen, Niansheng and Cheng, Songlin and Fan, Guangyu and Rao, Lei and Song, Xiaoyong and Lv, Wenjing and Yang, Dingyu},
  journal={Wirel. Net.},
  volume={30},
  number={5},
  pages={3061--3073},
  year={2024},
  publisher={Springer}
}

@article{cheng2025secrecy,
  title={{Secrecy analysis and optimization of UAV-assisted relaying networks with hardware impairments}},
  author={Cheng, Songlin and Chen, Jielin and Hu, Xiu and Chen, Niansheng and Fan, Guangyu and Rao, Lei and Song, Xiaoyong and Yang, Dingyu},
  journal={Electron. Lett.},
  volume={61},
  number={1},
  pages={e70248},
  year={2025},
  publisher={Wiley Online Library}
}

@ARTICLE{10510457,
  author={Alsmadi, Hamzih and Saleh, Emad and Alsmadi, Malek and Ikki, Salama},
  journal={IEEE Commun. Lett.}, 
  title={{Hardware impairments effects on over the air system assisted by unmanned aerial vehicle}}, 
  year={2024},
  volume={28},
  number={7},
  pages={1609-1613},
  doi={10.1109/LCOMM.2024.3395439}
}

@ARTICLE{10146460,
  author={Pandey, Gaurav K. and Gurjar, Devendra S. and Yadav, Suneel and Solanki, Sourabh},
  journal={IEEE Sens. Lett.}, 
  title={{UAV-empowered IoT network With hardware impairments and shadowing}}, 
  year={2023},
  volume={7},
  number={7},
  pages={1-4},
  doi={10.1109/LSENS.2023.3284089}
}

@ARTICLE{10287354,
  author={Minh, Bui Vu and Le, Anh-Tu and Le, Chi-Bao and Nguyen, Sang Quang and Phan, Van-Duc and Nguyen, Tan N. and Voznak, Miroslav},
  journal={IEEE Access}, 
  title={{Performance prediction in UAV-terrestrial networks with hardware noise}}, 
  year={2023},
  volume={11},
  number={},
  pages={117562-117575},
  doi={10.1109/ACCESS.2023.3325478}
}

@ARTICLE{10411132,
  author={Feng, Jihan and Liu, Xin and Liu, Zechen and Durrani, Tariq S.},
  journal={IEEE Trans. Veh. Tech.}, 
  title={{Optimal trajectory and resource allocation for RSMA-UAV assisted IoT communications}}, 
  year={2024},
  volume={73},
  number={6},
  pages={8693-8704},
  doi={10.1109/TVT.2024.3354329}
}

@article{yao2024coordinated,
  title={{Coordinated RSMA for integrated sensing and communication in emergency UAV systems}},
  author={Yao, Binghan and Li, Ruoguang and Chen, Yingyang and Wang, Li},
  journal={arXiv preprint arXiv:2406.19205},
  year={2024}
}

@article{hua2024sum,
  title={{On sum-rate maximization in downlink UAV-aided RSMA systems}},
  author={Hua, Duc-Thien and Do, Quang Tuan and Dao, Nhu-Ngoc and Cho, Sungrae},
  journal={ICT Express},
  volume={10},
  number={1},
  pages={15--21},
  year={2024},
  publisher={Elsevier}
}

@ARTICLE{11029408,
  author={Khennoufa, Faical and Abdellatif, Khelil and Yanikomeroglu, Halim and Ozturk, Metin and Elganimi, Taissir and Kara, Ferdi and Rabie, Khaled},
  journal={IEEE Internet of Things Mag.}, 
  title={{A multi-layer non-terrestrial networks architecture for 6G and beyond under realistic conditions and with practical limitations}}, 
  year={2025},
  volume={8},
  number={5},
  pages={136-143},
  doi={10.1109/MIOT.2025.3575923}
}

@ARTICLE{9519666,
  author={Singh, Sandeep Kumar and Agrawal, Kamal and Singh, Keshav and Li, Chih-Peng},
  journal={IEEE Wirel. Commun. Lett.}, 
  title={{Outage probability and throughput analysis of UAV-assisted rate-splitting multiple access}}, 
  year={2021},
  volume={10},
  number={11},
  pages={2528-2532},
  doi={10.1109/LWC.2021.3106456}
}

@ARTICLE{10188818,
  author={Khennoufa, Faical and Khelil, Abdellatif and Beddiaf, Safia and Kara, Ferdi and Rabie, Khaled and Kaya, Hakan and Emir, Ahmet and Ikki, Salama and Yanikomeroglu, Halim},
  journal={IEEE Access}, 
  title={{Wireless powered cooperative communication network for dual-hop uplink NOMA with IQI and SIC imperfections}}, 
  year={2023},
  volume={11},
  number={},
  pages={76506-76523},
  doi={10.1109/ACCESS.2023.3297487}
}

@inproceedings{lorincz2021novel,
  title={{A novel real-time unmanned aerial vehicles-based disaster management framework}},
  author={Lorincz, Josip and Tahirovi{\'c}, Adnan and Stojkoska, Biljana Risteska},
  booktitle={IEEE 29th Telecommun. Forum},
  pages={1--4},
  year={2021},
}

@article{kurt2021vision,
  title={{A vision and framework for the high altitude platform station (HAPS) networks of the future}},
  author={Kurt, Gunes Karabulut and Khoshkholgh, Mohammad G and Alfattani, Safwan and Ibrahim, Ahmed and Darwish, Tasneem SJ and Alam, Md Sahabul and Yanikomeroglu, Halim and Yongacoglu, Abbas},
  journal={IEEE Commun. Surv. Tutor.},
  volume={23},
  number={2},
  pages={729--779},
  year={2021},
  publisher={IEEE}
}

@article{trespalacios2015improved,
  title={{Improved Big-M reformulation for generalized disjunctive programs}},
  author={Trespalacios, Francisco and Grossmann, Ignacio E},
  journal={Comput. Chem. Eng.},
  volume={76},
  pages={98--103},
  year={2015},
  publisher={Elsevier}
}

@book{wolsey1999integer,
  title={{Integer and Combinatorial Optimization}},
  author={Wolsey, Laurence A and Nemhauser, George L},
  year={1999},
  publisher={John Wiley \& Sons}
}

@book{nocedal2006numerical,
  title={{Numerical optimization}},
  author={Nocedal, Jorge and Wright, Stephen J},
  year={2006},
  publisher={Springer}
}

@article{al2014optimal,
  title={{Optimal LAP altitude for maximum coverage}},
  author={Al-Hourani, Akram and Kandeepan, Sithamparanathan and Lardner, Simon},
  journal={IEEE Wirel. Commun. Lett.},
  volume={3},
  number={6},
  pages={569--572},
  year={2014},
  publisher={IEEE}
}

@article{khennoufa2024error,
  title={{Error performance analysis of UAV-mounted RIS for NOMA systems with practical constraints}},
  author={Khennoufa, Faical and Abdellatif, Khelil and Kara, Ferdi and Yanikomeroglu, Halim and Rabie, Khaled and Elganimi, Taissir Y and Beddiaf, Safia},
  journal={IEEE Commun. Lett.},
  volume={28},
  number={4},
  pages={887--891},
  year={2024},
  publisher={IEEE}
}

@inproceedings{mozaffari2015drone,
  title={{Drone small cells in the clouds: Design, deployment and performance analysis}},
  author={Mozaffari, Mohammad and Saad, Walid and Bennis, Mehdi and Debbah, Merouane},
  booktitle={ IEEE Global Commun. Conf. },
  pages={1--6},
  year={2015},
}

@article{beddiaf2023impact,
  title={{Impact of hardware impairment on the uplink SIMO Cooperative NOMA with selection relay under imperfect CSI}},
  author={Beddiaf, Safia and Khelil, Abdellatif and Khennoufa, Faical and Kara, Ferdi and Rabie, Khaled and Li, Xingwang and Kaya, Hakan and Emir, Ahmet and Yanikomeroglu, Halim},
  journal={IEEE Access},
  volume={11},
  pages={106706--106721},
  year={2023},
  publisher={IEEE}
}

@ARTICLE{10122143,
  author={Xiao, Meng and Cui, Huanxi and Huang, Dianrun and Zhao, Zhongliang and Cao, Xianbin and Wu, Dapeng Oliver},
  journal={IEEE Trans. Netw. Sci. Eng.}, 
  title={{Traffic-aware energy-efficient resource allocation for RSMA based UAV communications}}, 
  year={2024},
  volume={11},
  number={3},
  pages={2537-2548},
  doi={10.1109/TNSE.2023.3274550}
}

@ARTICLE{9258414,
  author={Jaafar, Wael and Naser, Shimaa and Muhaidat, Sami and Sofotasios, Paschalis C. and Yanikomeroglu, Halim},
  journal={IEEE Trans. Veh. Tech.}, 
  title={{On the downlink performance of RSMA-based UAV communications}}, 
  year={2020},
  volume={69},
  number={12},
  pages={16258-16263},
  doi={10.1109/TVT.2020.3037657}
}

@article{babaei2018ber,
  title={{BER analysis of dual-hop relaying with energy harvesting in nakagami-$ m $ fading channel}},
  author={Babaei, Mohammadreza and Ayg{\"o}l{\"u}, {\"U}mit and Basar, Ertugrul},
  journal={IEEE Trans. Wirel. Commun.},
  volume={17},
  number={7},
  pages={4352--4361},
  year={2018},
  publisher={IEEE}
}

@article{babaei2022performance,
  title={{Performance analysis of dual-hop AF relaying with non-linear/linear energy harvesting}},
  author={Babaei, Mohammadreza and Durak-Ata, L{\"u}tfiye and Ayg{\"o}l{\"u}, {\"U}mit},
  journal={Sensors},
  volume={22},
  number={16},
  pages={5987},
  year={2022},
  publisher={MDPI}
}

@book{pardo2020statistical,
  title={{Statistical Analysis of Empirical Data}},
  author={Pardo, Scott},
  year={2020},
  publisher={Springer}
}

@article{arzykulov2021hardware,
  title={Hardware and interference limited cooperative CR-NOMA networks under imperfect SIC and CSI},
  author={Arzykulov, Sultangali and Nauryzbayev, Galymzhan and Celik, Abdulkadir and Eltawil, Ahmed M},
  journal={IEEE Open J. Commun. Soc.},
  volume={2},
  pages={1473--1485},
  year={2021},
  publisher={IEEE}
}

@article{morales1989generalization,
  title={The generalization of the binomial theorem},
  author={Morales, J and Flores-Riveros, A},
  journal={J. Math. Phys.},
  volume={30},
  number={2},
  pages={393--397},
  year={1989},
  publisher={American Institute of Physics}
}

@misc{thomas2009introduction,
  title={{Introduction to algorithms third edition}},
  author={Thomas H, Cormen and Charles, E and Ronald L, Rivest and Clifford, Stein and others},
  year={2009},
  publisher={Mit Press}
}

@article{hoang2020outage,
  title={{Outage and throughput analysis of power-beacon assisted nonlinear energy harvesting NOMA multi-user relay system over Nakagami-m fading channels}},
  author={Hoang, Tran Manh and Nguyen, Ba Cao and Trung, Tran Thanh and others},
  journal={Heliyon},
  volume={6},
  number={11},
  year={2020},
  publisher={Elsevier}
}

@article{li2025outage,
  title={{Outage probability for downlink multi-antenna RSMA system with imperfect SIC}},
  author={Li, Shenhong and Zhu, Juan and Feng, Tao and Derakhshani, Mahsa},
  journal={IEEE Wirel. Commun. Lett.},
  volume={15},
  pages={385--389},
  year={2025},
  publisher={IEEE}
}

@ARTICLE{10430407,
  author={Qin, Langtian and Lu, Hancheng and Chen, Yuang and Chong, Baolin and Guo, Fengqian},
  journal={IEEE Trans. Veh. Tech.}, 
  title={{Joint rransmission and resource optimization in NOMA-assisted IoVT with mobile edge computing}}, 
  year={2024},
  volume={73},
  number={7},
  pages={9984-9999},
  doi={10.1109/TVT.2024.3364358}
  }

@ARTICLE{11037391,
  author={Chen, Yuang and Lu, Hancheng and Wu, Chang and Qin, Langtian and Guo, Xiaobo},
  journal={IEEE Trans. Wirel. Commun.}, 
  title={{Performance optimization in RSMA-assisted uplink xURLLC IIoT networks with statistical QoS provisioning}}, 
  year={2025},
  volume={24},
  number={12},
  pages={10100-10115},
  doi={10.1109/TWC.2025.3577694}
  }

@article{mao2022rate,
  title={{Rate-splitting multiple access: Fundamentals, survey, and future research trends}},
  author={Mao, Yijie and Dizdar, Onur and Clerckx, Bruno and Schober, Robert and Popovski, Petar and Poor, H Vincent},
  journal={IEEE Commun. Surv. Tutor.},
  volume={24},
  number={4},
  pages={2073--2126},
  year={2022},
  publisher={IEEE}
}

@ARTICLE{10851439,
  author={Liu, Heyou and Bashir, Muhammad Salman and Alouini, Mohamed-Slim},
  journal={IEEE Trans. Aerosp. Electron. Syst.}, 
  title={{3-D position optimization of solar-powered hovering UAV relay in optical wireless backhaul}}, 
  year={2025},
  volume={61},
  number={3},
  pages={5853-5870},
  doi={10.1109/TAES.2025.3526742}}

@ARTICLE{11441975,
  author={Pei, Xinyue and Liu, Jihao and Luo, Xuewen and Wang, Xingwei and Chen, Yingyang and Wen, Miaowen and Tsiftsis, Theodoros A.},
  journal={IEEE Trans. Veh. Technol.}, 
  title={{RSMA-enabled covert communications against multiple spatially random wardens}}, 
  year={2026},
  volume={},
  number={},
  pages={1-6},
  doi={10.1109/TVT.2026.3675333}}

@ARTICLE{11370792,
  author={Wei, Xinyi and Li, Ruoguang and Chen, Yingyang and Xu, Lianming and Wang, Li and Han, Zhu},
  journal={IEEE Trans. Cogn. Commun. Netw.}, 
  title={{Coordinated rate-splitting multiple access for emergency UAV-enabled integrated sensing and communication}}, 
  year={2026},
  volume={12},
  number={},
  pages={5999-6015},
  doi={10.1109/TCCN.2026.3660777}}

@ARTICLE{11159579,
  author={Wang, Yi and Chen, Yingyang and Wang, Li and Cai, Donghong and Li, Xiaofan and Fan, Pingzhi},
  journal={IEEE Trans. Wirel. Commun.}, 
  title={{Autonomous driving with RSMA-enabled finite blocklength transmissions: ergodic performance analysis and optimization}}, 
  year={2026},
  volume={25},
  number={},
  pages={2954-2969},
  doi={10.1109/TWC.2025.3600546}}

\begin{IEEEbiographynophoto}{Faicel Khennoufa} is an Assistant Professor of Telecommunications at the École Nationale Supérieure des Technologies Avancées (ENSTA), and a member of the Innovative Technologies Laboratory (LTI), ENSTA, Algiers, Algeria. His research interests include wireless communication systems and non-terrestrial networks.
\end{IEEEbiographynophoto}

\begin{IEEEbiographynophoto}{Abdellatif Khelil} is a Full Professor of Communication Engineering at El-Oued University, Algeria. His research focuses on wireless communications, cellular communications (5G, B5G, 6G), MIMO systems, mm-wave propagation, THz communications, new waveforms, NOMA, RSMA, RIS, and NTN communications.
\end{IEEEbiographynophoto}

\begin{IEEEbiographynophoto}{Metin Ozturk (Senior Member, IEEE)} is an Assistant Professor of Telecommunications at Ankara Yıldırım Beyazıt University, Türkiye. His research interests include wireless communications, with a particular focus on AI-driven mobile networking, non-terrestrial networks, and sustainable wireless network design. 
\end{IEEEbiographynophoto}

\begin{IEEEbiographynophoto}{Halim Yanikomeroglu (Fellow, IEEE)} is a Chancellor’s Professor in the Department of Systems and Computer Engineering at Carleton University, Canada, and the Director of Carleton-NTN (Non-Terrestrial Networks) Lab. His research interests span the overall systems architecture of wireless networks, as well as the physical, medium access, and networking layers, and the cross-layer aspects of wireless communications. 
\end{IEEEbiographynophoto}

\begin{IEEEbiographynophoto}{Safwan Alfattani (Member, IEEE)} is an Assistant Professor in the Electrical Engineering Department-Rabigh at King Abdulaziz University, Saudi Arabia. His research interests include wireless communications, non-terrestrial networks, reconfigurable intelligent surfaces, and IoT networks.
\end{IEEEbiographynophoto}

\vfill

\end{document}